\documentclass{aa}
\usepackage[latin1]{inputenc}
\usepackage{graphics,psfig}
\usepackage{amssymb}
\usepackage{mathrsfs}
\usepackage{natbib}
\bibpunct{(}{)}{;}{a}{}{,}

\newcommand\kms{\ensuremath{\mathrm{km}\,\mathrm{s}^{-1}}}

\newcommand\Amm{\AA\,$\mathrm{mm}^{-1}$}
\newcommand\Teff{\ensuremath{T_\mathrm{eff}}}
\newcommand\logg{\ensuremath{\log g}}
\newcommand\vsini{\ensuremath{v\sin i}}

\begin{document}

\title{Rotational velocities of A-type stars \thanks{Based on
    observations made at Observatoire de Haute Provence (CNRS),
    France}\fnmsep\thanks{Tables \ref{results} and \ref{merging} are only available in electronic form at the CDS via anonymous ftp to cdsarc.u-strasbg.fr (130.79.125.5) or via http://cdsweb.u-strasbg.fr/Abstract.html}}

\subtitle{II. Measurement of \vsini\ in the northern hemisphere}

\author{F. Royer\inst{1,2} 
  \and S. Grenier\inst{2}
  \and M.-O. Baylac\inst{2} 
  \and A.E. G\'omez\inst{2}  
  \and J. Zorec\inst{3}
}

\offprints{Frédéric Royer}
 
\institute{Observatoire de Gen\`eve, 51 chemin des Maillettes, CH-1290 Sauverny, Switzerland
      \and GEPI/CNRS FRE 2459, Observatoire de Paris, 5 place Janssen, F-92195 Meudon cedex, France
      \and CNRS, Institut d'Astrophysique de Paris, 98 bis boulevard Arago, F-75014 Paris, France}

\date{Received / Accepted}

\titlerunning{Rotational velocities of A-type stars. II.}

\abstract{This work is the second part of the set of measurements of
  \vsini\ for A-type stars, begun by \citet{Ror_02a}. Spectra of 249 B8 to F2-type stars
  brighter than $V=7$ have been collected at Observatoire de
  Haute-Provence (OHP). Fourier transforms of several line profiles in the range 4200--4600~\AA\ are used to derive \vsini\ from the frequency of the first zero. Statistical analysis of the sample indicates that measurement error mainly depends on \vsini\ and this relative error of the rotational velocity is found to be about 5~\% on average.\\
The systematic shift with respect to standard values from \citet{Slk_75}, previously
found in the first paper, is here confirmed. Comparisons with data
from the literature agree with our findings: \vsini\ values from
Slettebak et al. are underestimated and the relation between both
scales follows a linear law $\vsini_\mathrm{new} = 1.03\,\vsini_\mathrm{old}+7.7$. \\
Finally, these data are combined with those from the previous paper
\citep{Ror_02a}, together with the catalogue of \citet{AbtMol95}. The
resulting sample includes some 2150 stars with homogenized rotational velocities.
\keywords{techniques: spectroscopic -- stars: early-type; rotation}}

\maketitle

\section{Introduction}
This paper is a continuation of the rotational velocity study of
A-type stars, initiated in \citet[ hereafter Paper~I]{Ror_02a}. The main goals and
motivations are described in the previous paper. The sample of A-type stars
described and analyzed in this work is the counterpart of the one in
Paper~I, in the northern hemisphere.

In short, it is intended to produce a homogeneous sample of measurements of
projected rotational velocities (\vsini) 
for the spectral interval of A-type  stars, and this without using any 
preset calibration. 
                            
This article is structured in a way identical to the precedent, except for
an additional section (Sect.~\ref{merge}) where  data from this paper, the previous one and the catalogue of
\citet{AbtMol95} are gathered, and the total sample is discussed in statistical terms. 
  
\section{Observational data}

Spectra were obtained in the northern hemisphere with the AUR\'ELIE
spectrograph \citep{Git_94} associated with the 1.52~m telescope at
Observatoire de Haute-Provence (OHP), in order to acquire
complementary data to HIPPARCOS observations \citep{GrrBue95}.

\begin{figure}[!htp]
  \centering \resizebox{\hsize}{!}{\includegraphics{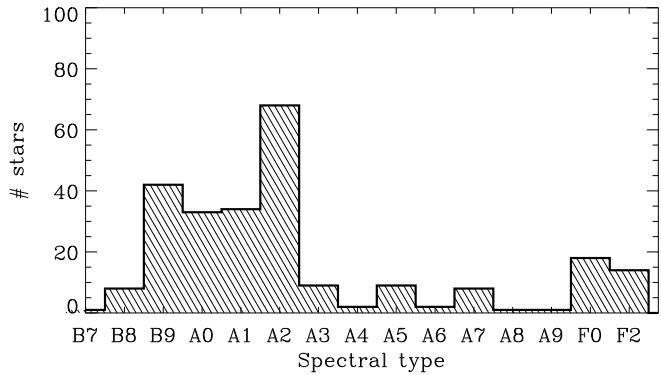}}
        \caption{Distribution of the spectral type for the 249 programme stars.}
        \label{histST}
\end{figure}

The initial programme gathers early-type stars for which \vsini\ 
measurement is needed.  More than 820 spectra have been collected for
249 early-type stars from January 1991 to May 1994.  As shown in
Fig.~\ref{histST}, B9 to A2-type stars represent the major part of the
sample (70~\%). Most of the stars are on the main sequence and only
about one fourth are classified as more evolved than the luminosity
class III-IV.

These northern stars are brighter than the magnitude $V=7$. Nevertheless, three
stars are fainter than this limit and do not belong to the
HIPPARCOS Catalogue \citep{Hip}. Derivation of their magnitude from TYCHO observations
turned out to be: HD~23643 $V=7.79$, HD~73576 $V=7.65$ and HD~73763 $V=7.80$. These 
additional stars are special targets known to be $\delta$~Scuti stars.


\begin{figure*}[!htp]
  \centering \resizebox{12cm}{!}{\includegraphics{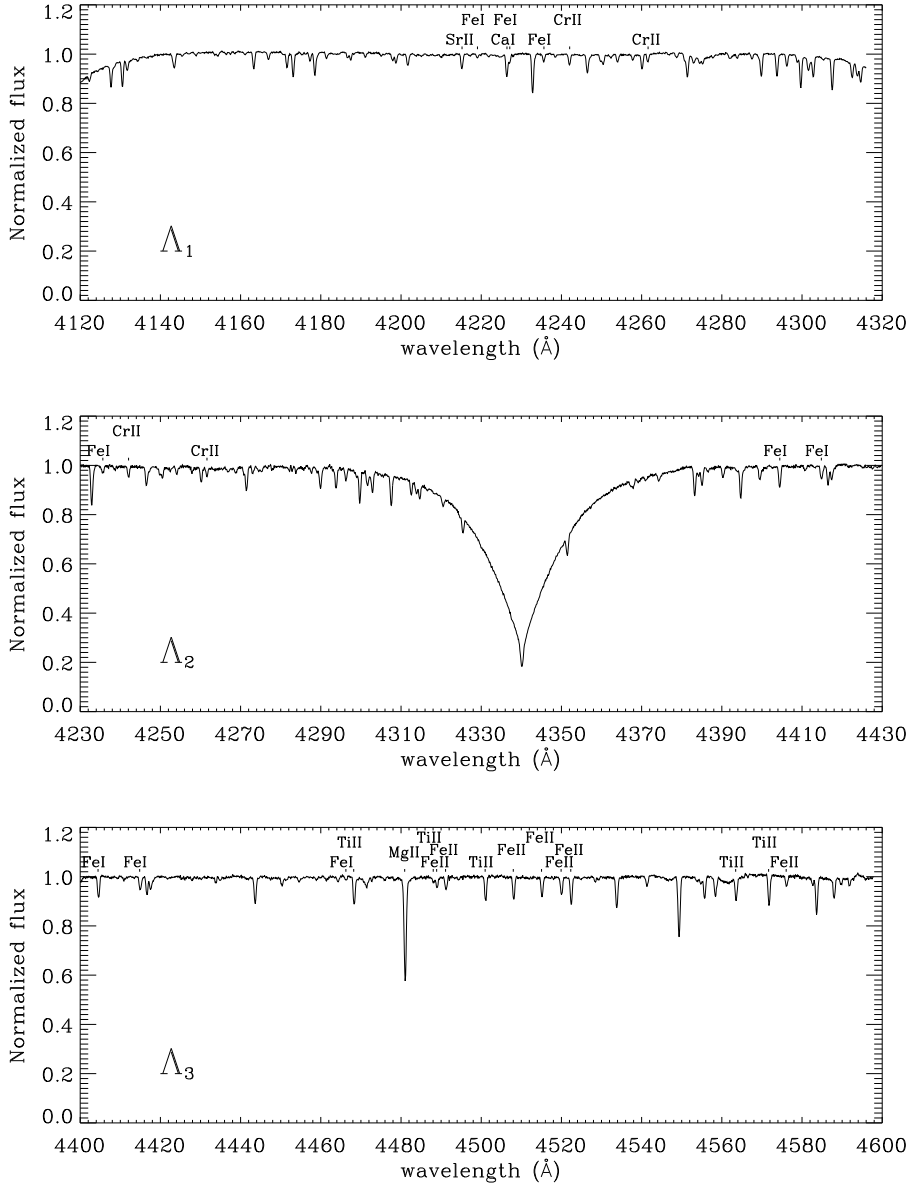}}
  \hfill \parbox[b]{55mm}{
\caption{Observed spectra of Vega are displayed for the different spectral ranges: top panel, range $\Lambda_1$; middle panel, range $\Lambda_2$; bottom panel, range $\Lambda_3$. Each domain covers nearly 200~\AA.}
 \begin{itemize} \item $\Lambda_1$: the first range extends from the
   red wing of H$\delta$ to the blue wing of H$\gamma$ which restricts
   the reliable normalization area. It contains seven of the selected
   lines. \item $\Lambda_2$: centered around H$\gamma$, this range
   only contains five selected lines. \item $\Lambda_3$: this range
   contains the largest number of selected lines, 16 in total, among
   which the doublet line \ion{Mg}{ii} 4481. \end{itemize} The 23
 selected lines (listed in Table~\ref{raies_vsini}) are indicated, and
 show up twice in  the overlap areas.The instrumental feature coming from flat-fielding lamp is noticeable in the three spectra ($\sim 4280$~\AA\ in $\Lambda_1$, $\sim 4390$~\AA\ in $\Lambda_2$, $\sim 4560$~\AA\ in $\Lambda_3$).
        \label{spectre}
        }
\end{figure*}

AUR\'ELIE spectra were obtained in three different spectral ranges
(Fig.~\ref{spectre}):
	\begin{itemize}
	\item $\Lambda_1$: 4110--4310~\AA. This domain was initially
          chosen to detect chemically peculiar stars by measuring
          equivalent widths of the
          lines \ion{Si}{ii} 4128, \ion{Si}{ii} 4131, \ion{Sr}{ii}
            4216, and \ion{Sc}{ii} 4247.  
	\item $\Lambda_2$: 4226--4432~\AA\ or 4270--4475~\AA. Centered 
          around H$\gamma$, this spectral range allows the
          determination of the effective temperature.
        \item $\Lambda_3$: 4390--4600~\AA, or 4400--4610~\AA. This
          domain is dedicated to \vsini\ determination,
          although all the three ranges are used to measure the
          rotational velocity.
	\end{itemize}
        
        Two thirds of the sample have observations in each of the three
        ranges.  
        The $\Lambda_3$ range is particularly aimed at \vsini\
        measurement, and  contains the largest number
        of lines selected for this purpose (twice as much
        as $\Lambda_1$ and $\Lambda_2$). Besides it is the only one which
        covers the magnesium doublet at 4481~\AA. This line remains often
        alone for measurement in fast rotators. It is thus significant
        to note that among the 249 stars of the sample, three were
        observed only in the $\Lambda_1$ range, eight only in
        $\Lambda_2$ and eleven in both $\Lambda_1$ and $\Lambda_2$
        only. Overall 22 stars have no observation in
        $\Lambda_3$. The reason is that these stars have already
        known \vsini. Moreover they are effective temperature standard
        stars or reference stars for chemical abundances.
        
        Some changes in the configuration of the instrument meant the
        central wavelengths of the $\Lambda_2$ and $\Lambda_3$ domains
        have been slightly modified during mission period.
                
        The entrance of the slit is a 600~$\mu$m hole, i.e. 3\arcsec\ 
        on the sky, dedicated to the 1.52~m Coudé telescope. The
        dispersion of the collected spectra is 8.1 ~\Amm\ and the
        resolving power is about 16\,000.
        
        The barrette detector is a double linear array TH~7832, made
        of 2048 photo-diodes.  Reduction of the data has been processed
        using MIDAS\footnote{MIDAS is being developed and maintained
          by ESO} procedures.
        
        Flat field correction with a tungsten lamp and wavelength
        calibration with a Th-Ar lamp have been made with classical
        procedures.  Nevertheless, a problem occurred when applying the
        flat-field correction with the tungsten calibration lamp. The
        division by the W lamp spectrum produced a spurious effect in
        the resulting spectrum at a given position in the pixels axis.
        This effect distorts the continuum, as it can be seen on the
        spectra of Vega (Fig.~\ref{spectre}). This problem triggered
        the decision to change the instrumental configuration and central wavelengths
        of the spectral ranges. 

\section{Measurement of the rotational velocity}

The method adopted for \vsini\ determination is the computation of
the first zero of Fourier transform (FT) of line profiles
\citep{Cal33,Raa_89}. For further description of the method applied
to our sample, see Paper~I.
The different observed spectral range induces some changes, 
which are detailed below.

\subsection{Continuum tracing}
The normalization of the spectra was performed using MIDAS: the continuum has been determined
visually, passing through noise fluctuations. The procedure is much
like the normalization carried out in Paper~I, except for a different spectral window. 
For the ranges $\Lambda_1$ and $\Lambda_2$, the influence of the
Balmer lines is important, and their wings act as non negligible
contributions to the difference between true and pseudo-continuum, over
the major part of the spectral domain, as shown in Paper~I.
On the other hand, the $\Lambda_3$ range is further from H$\gamma$.
In order to quantify the alteration of continuum due to Balmer lines
wings and blends of spectral lines, a grid of synthetic spectra of
different effective temperatures (10\,000, 9200, 8500 and 7500\,K) and
different rotational broadenings, computed from Kurucz' model
atmosphere \citep{Kuz93}, is used to calculate the differences between
the true continuum and the pseudo-continuum. The pseudo-continuum is
represented as the highest points in the spectra. The differences
are listed in Table~\ref{continuum}, for different spectral 20~\AA\
wide sub-ranges. This table is a continuation of the similar one in
Paper~I, considering the spectral range 4200--4500~\AA. 
\begin{table}[!htp]
\begin{center}
\caption{Differences between the true continuum and the highest points
  in different spectral bands for the set of synthetic
  spectra in the $\Lambda_3$ domain. Wavelength indicates the center of the 20~\AA\
wide range.}
\label{continuum}
\setlength\tabcolsep{5pt}
\begin{tabular}{rrccccc}
\hline
\multicolumn{2}{c}{\Teff, \vsini} & \multicolumn{5}{c}{central wavelength (\AA)} \cr
\multicolumn{2}{c}{(K, \kms)} & 4510 & 4530 & 4550 & 4570 & 4590  \cr
\hline
\multicolumn{7}{c}{\sl Data for wavelengths shorter than 4500~\AA}\cr
\multicolumn{7}{c}{\sl are given in Table~1 of Paper I}\cr
\hline
10\,000,&  10 & 0.0005 & 0.0003 & 0.0002 & 0.0000 & 0.0000 \cr
10\,000,&  50 & 0.0008 & 0.0003 & 0.0003 & 0.0002 & 0.0003 \cr
10\,000,& 100 & 0.0011 & 0.0005 & 0.0016 & 0.0005 & 0.0013 \cr

   9200,&  10 & 0.0010 & 0.0006 & 0.0006 & 0.0006 & 0.0006 \cr
   9200,&  50 & 0.0017 & 0.0008 & 0.0010 & 0.0012 & 0.0012 \cr
   9200,& 100 & 0.0023 & 0.0012 & 0.0027 & 0.0012 & 0.0051 \cr

   8500,&  10 & 0.0017 & 0.0012 & 0.0010 & 0.0010 & 0.0010 \cr
   8500,&  50 & 0.0030 & 0.0020 & 0.0022 & 0.0025 & 0.0020 \cr
   8500,& 100 & 0.0042 & 0.0027 & 0.0062 & 0.0030 & 0.0093 \cr

   7500,&  10 & 0.0005 & 0.0005 & 0.0005 & 0.0005 & 0.0005 \cr
   7500,&  50 & 0.0032 & 0.0023 & 0.0036 & 0.0045 & 0.0032 \cr
   7500,& 100 & 0.0059 & 0.0050 & 0.0149 & 0.0059 & 0.0181 \cr
\hline
\end{tabular}
\end{center}
\end{table}
 It is clear that the pseudo-continuum is much closer to the true
 continuum in $\Lambda_3$ than in both bluer ranges. 

\subsection{Set of lines} 
  
Put end to end, the spectra acquired with AUR\'ELIE cover a spectral
range of almost 500 Å. It includes that observed with ECHELEC in
Paper~I. The choice of the lines for the determination of the \vsini\ 
in Paper~I is thus still valid here. Moreover, in addition to this
selection, were adopted redder lines in order to benefit from the
larger spectral coverage.

The complete list of the 23 lines that are candidate for \vsini\ 
determination is given in Table~\ref{raies_vsini}.

\begin{table}[!htp] 
\begin{center} 
\caption{List of the 23 spectral lines used (when possible) for the
  \vsini\ measurement, and the corresponding spectral range(s) to
  which they belong.}
\label{raies_vsini}
\begin{tabular}{c|ll|c} 
\hline 
range & wavelength & element & range \cr 
\hline 
            & 4215.519 & \ion{Sr}{ii} \cr
            & 4219.360 & \ion{Fe}{i}  \cr
            & 4226.728 & \ion{Ca}{i}  \cr
$\Lambda_1$ & 4227.426 & \ion{Fe}{i}  \cr
                                    \cline{2-4}  
            & 4235.936 & \ion{Fe}{i}  \cr
            & 4242.364 & \ion{Cr}{ii} \cr
            & 4261.913 & \ion{Cr}{ii} & $\Lambda_2$\cr
\cline{1-3} 
            & 4404.750 & \ion{Fe}{i}  \cr
            & 4415.122 & \ion{Fe}{i}  \cr
                                    \cline{2-4}  
            & 4466.551 & \ion{Fe}{i}  \cr
            & 4468.507 & \ion{Ti}{ii} \cr
            & 4481$^{.126} _{.325}$ & \ion{Mg}{ii} \dag \cr
            & 4488.331 & \ion{Ti}{ii} \cr
$\Lambda_3$ & 4489.183 & \ion{Fe}{ii} \cr
            & 4491.405 & \ion{Fe}{ii} \cr
            & 4501.273 & \ion{Ti}{ii} \cr
            & 4508.288 & \ion{Fe}{ii} \cr
            & 4515.339 & \ion{Fe}{ii} \cr
            & 4520.224 & \ion{Fe}{ii} \cr
            & 4522.634 & \ion{Fe}{ii} \cr
            & 4563.761 & \ion{Ti}{ii} \cr
            & 4571.968 & \ion{Ti}{ii} \cr
            & 4576.340 & \ion{Fe}{ii} \cr

\hline
\end{tabular}
\end{center}
\dag\ Wavelength of both components are indicated for the magnesium doublet line.
\end{table}

In order to quantify effects of blends in the selected lines for later
spectral types, we use the skewness of synthetic line profiles, as in
Paper~I. The same grid of synthetic spectra computed using Kurucz'
model \citep{Kuz93}, is used.  Skewness is defined as $\gamma_1 = m_3
\, m_2^{-1.5}$, where $m_k$ is moment of $k$-th order equal to

\begin{equation} 
\label{moment}
\forall k,\;m_k = {\displaystyle \sum_{i=1}^{L}\left[1-\mathscr{F}(\lambda_i)\right]  \left[\lambda_i-\lambda_\mathrm{c}\right]^k 
 \over \displaystyle \sum_{i=1}^{L} 1-\mathscr{F}(\lambda_i) },
\end{equation} for an absorption line centered at wavelength $\lambda_\mathrm{c}$ and spreading from $\lambda_1$ to $\lambda_L$, where $\mathscr{F}(\lambda_i)$ is the normalized flux corresponding to the wavelength $\lambda_i$. Ranges $[\lambda_1,\lambda_L]$ are centered around theoretical wavelengths from Table~\ref{raies_vsini} and the width of the window is taken to be 0.35, 0.90 and 1.80~\AA\ for rotational broadening 10, 50 and 100~\kms\ respectively (the width around the \ion{Mg}{ii} doublet is larger: 1.40, 2.0 and 2.3 \AA).
\begin{table}[!htp]
\begin{center}
\caption{Variation of the skewness $\gamma_1$ of the lines with \Teff\ and \vsini\ in the synthetic spectra.}
\label{skewness}
\begin{tabular}{lrrrrr}
\hline
 & \multicolumn{1}{c}{\vsini} & \multicolumn{4}{c}{\Teff\ (K)} \cr
\cline{3-6}
line& \multicolumn{1}{c}{(\kms)} &10\,000 &        9200   &         8500  &         7500 \cr
\hline   
\multicolumn{6}{c}{\sl Data for wavelengths shorter than 4500~\AA}\cr
\multicolumn{6}{c}{\sl are given in Table~3 of Paper I}\cr
\hline

\ion{Ti}{ii} 4501 &  10~~~~ & $-0.05$ & $-0.06$ & $-0.07$ & $-0.12$ \\
                  &  50~~~~ & $-0.02$ & $-0.03$ & $-0.04$ & $-0.04$ \\
                  & 100~~~~ & $-0.03$ & $-0.04$ & $-0.05$ & $-0.07$ \\
\cline{2-6}
\ion{Fe}{ii} 4508 &  10~~~~ & $ 0.01$ & $ 0.01$ & $ 0.01$ & $ 0.02$ \\
                  &  50~~~~ & $-0.00$ & $-0.00$ & $-0.00$ & $-0.00$ \\
                  & 100~~~~ & $-0.01$ & $-0.02$ & $-0.03$ & $-0.05$ \\
\cline{2-6}
\ion{Fe}{ii} 4515 &  10~~~~ & $ 0.00$ & $-0.00$ & $-0.01$ & $-0.06$ \\
                  &  50~~~~ & $ 0.02$ & $ 0.02$ & $ 0.01$ & $-0.04$ \\
                  & 100~~~~ & $ 0.01$ & $ 0.01$ & $ 0.02$ & $ 0.03$ \\
\cline{2-6}
\ion{Fe}{ii} 4520 &  10~~~~ & $ 0.01$ & $ 0.01$ & $ 0.01$ & $-0.01$ \\
                  &  50~~~~ & $ 0.00$ & $ 0.00$ & $-0.00$ & $-0.01$ \\
                  & 100~~~~ & $-0.17$ & $-0.19$ & $-0.23$ & $-0.30$ \\
\cline{2-6}
\ion{Fe}{ii} 4523 &  10~~~~ & $-0.06$ & $-0.06$ & $-0.06$ & $-0.05$ \\
                  &  50~~~~ & $-0.01$ & $-0.01$ & $-0.01$ & $ 0.01$ \\
                  & 100~~~~ & $-0.12$ & $-0.09$ & $-0.01$ & $ 0.08$ \\
\cline{2-6}
\ion{Ti}{ii} 4564 &  10~~~~ & $ 0.04$ & $ 0.04$ & $ 0.05$ & $ 0.06$ \\
                  &  50~~~~ & $ 0.01$ & $ 0.02$ & $ 0.04$ & $ 0.06$ \\
                  & 100~~~~ & $ 0.03$ & $ 0.04$ & $ 0.08$ & $ 0.16$ \\
\cline{2-6}
\ion{Ti}{ii} 4572 &  10~~~~ & $-0.00$ & $-0.00$ & $-0.01$ & $-0.02$ \\
                  &  50~~~~ & $ 0.01$ & $ 0.00$ & $-0.01$ & $-0.09$ \\
                  & 100~~~~ & $ 0.01$ & $ 0.01$ & $-0.00$ & $-0.04$ \\
\cline{2-6}
\ion{Fe}{ii} 4576 &  10~~~~ & $ 0.01$ & $ 0.01$ & $ 0.02$ & $ 0.05$ \\
                  &  50~~~~ & $ 0.00$ & $ 0.00$ & $ 0.01$ & $ 0.01$ \\
                  & 100~~~~ & $ 0.01$ & $ 0.02$ & $ 0.04$ & $ 0.07$ \\

\hline
\end{tabular}
\end{center}
\end{table}
Table~\ref{skewness} lists the skewness of the lines for each element of the synthetic spectra grid and is a continuation of
Table~3 from Paper~I for the lines with wavelength longer than
4500~\AA. These additional lines are rather isolated and free from
blends. Major part of the computed $\gamma_1$ for the hotter spectrum
(10\,000~K) is far lower than the threshold 0.15 chosen in Paper~I to
identify occurrence of blends. The only case where a line must be
discarded is the blend occurring with \ion{Fe}{ii} 4520 and
\ion{Fe}{ii} 4523 for $\vsini\gtrsim 100$~\kms. This non-blended
behavior continues on the whole range of temperature, and the
candidate lines remain reliable in most cases.
\begin{figure}[htp]
  \centering \resizebox{\hsize}{!}{\includegraphics{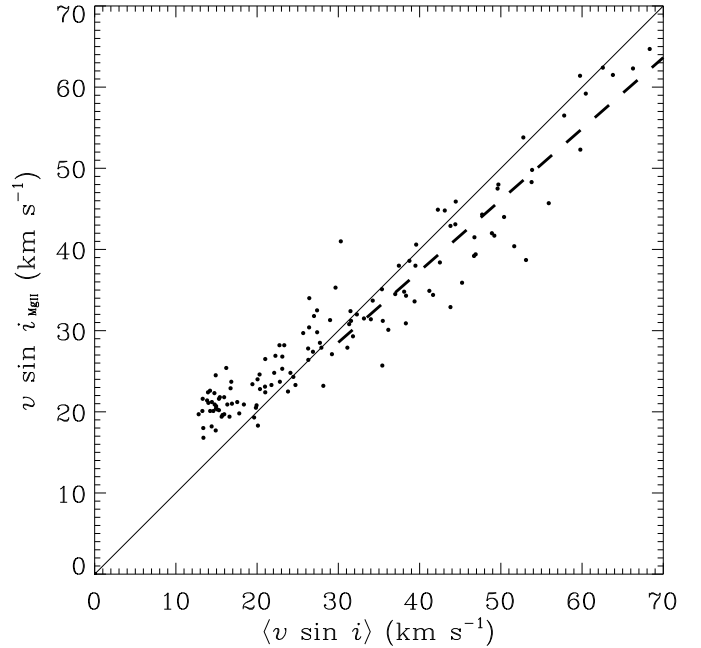}}
        \caption{$\vsini_\mathrm{\ion{Mg}{ii}}$ derived from the 4481
          \ion{Mg}{ii} line versus $\langle\vsini\rangle$ derived from
          other metallic lines for early A-type stars. The solid line
          stands for the one-to-one relation. The dashed line is the
          least-squares linear fit for
          $\langle\vsini\rangle>30$~\kms.}
        \label{vsini_MgII}
\end{figure}
\begin{figure}[htp]
  \centering \resizebox{\hsize}{!}{\includegraphics{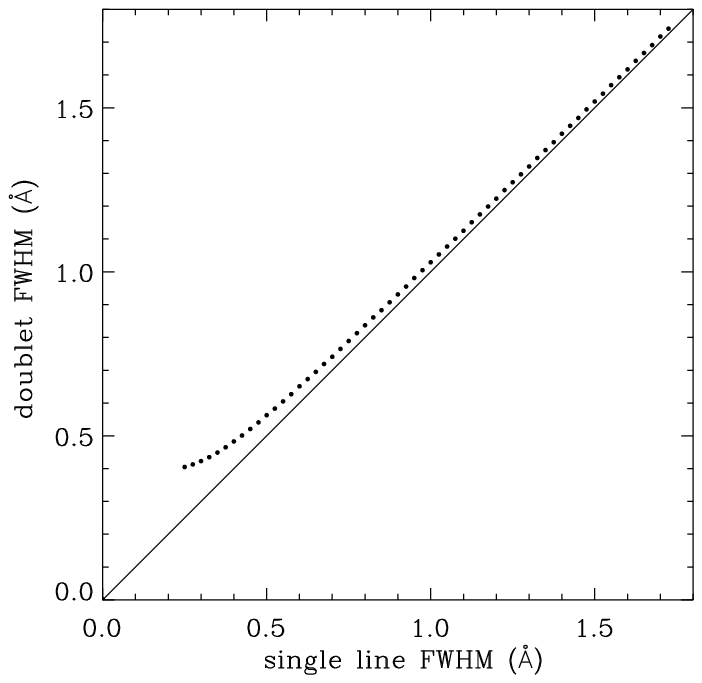}}
        \caption{Simulation of the doublet width behavior: $FWHM$ of the sum of two gaussian lines (separated with 
          0.2~\AA) as a function of the $FWHM$ of the components.}
        \label{vsini_MgII2}
\end{figure}

The comparison between the rotational velocity derived from the weak
lines and the one derived from the magnesium doublet was already
approached in Paper~I. It is here of an increased importance since the
\ion{Mg}{ii} line is not present in all spectra (i.e. $\Lambda_1$
and $\Lambda_2$ spectral ranges).  Figure~\ref{vsini_MgII} shows
this comparison between $\langle\vsini\rangle$ and
$\vsini_\mathrm{\ion{Mg}{ii}}$ using AUR\'ELIE data.  The deviation from the
one-to-one relation (solid line) in the low velocity part of the diagram
is due to the intrinsic width of the doublet. This deviation is
simulated by representing the \ion{Mg}{ii} doublet as the sum of two
identical gaussians separated by 0.2~\AA. The full-width at half
maximum ($FWHM$) of the simulated doublet line is plotted in
Fig.~\ref{vsini_MgII2} versus the $FWHM$ of its single-lined components.
The relation clearly deviates from the one-to-one relation for single
line $FWHM$ lower than 0.6~\AA. Using the rule of thumb from \citet[ hereafter SCBWP]{Slk_75}:
$FWHM{\scriptstyle [\mathrm{\AA}]} \approx
0.025\,\vsini{\scriptstyle [\kms]}$, this value corresponds to
$\vsini = 24$~\kms.  This limit coincides with what is observed in
Fig.~\ref{vsini_MgII}. For higher velocities
($\langle\vsini\rangle>30$~\kms), $\langle\vsini\rangle$ becomes
larger than {$\vsini_\mathrm{\ion{Mg}{ii}}$. A linear regression
  gives:
\begin{equation} 
\label{mgii}
\vsini_{\mathrm{\ion{Mg}{ii}}} = 0.88\,\langle\vsini\rangle +2.2.
\end{equation}
The effect is similar to the one found in Paper~I, suggesting that
blends in lines weaker than \ion{Mg}{ii} produce an overestimation of
the derived \vsini\ of about 10~\%.
\begin{figure}[!htp]
  \centering \resizebox{\hsize}{!}{\includegraphics{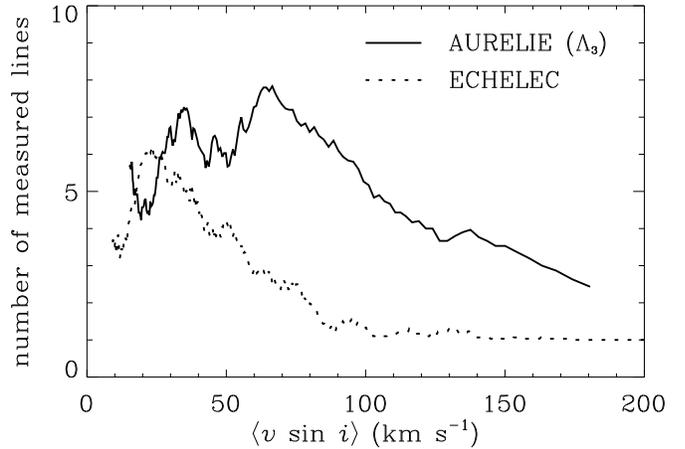}}
        \caption{The average number of measured lines (running average
          over 30 points) is plotted as a function of the mean
          $\langle\vsini\rangle$. Solid lines stands for the spectra
          collected with AUR\'ELIE ($\Lambda_3$ range) whereas dotted
          line represents ECHELEC spectra from Paper~I.}
        \label{line-vsini}
\end{figure}

The number of measurable lines among the 23 listed in
Table~\ref{raies_vsini} varies from one spectrum to another according
to the wavelength window, the rotational broadening and the
signal-to-noise ratio. The number of measured lines ranges from 1 to 17 lines. The
$\Lambda_3$ range offers a large number of candidate lines.
Fig.~\ref{line-vsini} shows the variation of this number with \vsini\
(solid line).
Rotational broadening starts to make the number of lines decrease
beyond about 70~\kms. Nevertheless additional lines in the spectral
domain redder than 4500~\AA\ makes the number of lines larger than in the
domain collected with ECHELEC (Paper~I; dotted line). Whereas with
ECHELEC the number of lines decreases with \vsini\ from 30~\kms\ to
reach only one line (i.e. the \ion{Mg}{ii} doublet) at 100~\kms, the number
of lines with AUR\'ELIE is much sizeable: seven at 70~\kms, still
four at 100~\kms\ and more than two even beyond 150~\kms.

\subsection{Precision} 
\label{precision}
\subsubsection{Effect of \vsini} 
In Fig.~\ref{sigma-vsini1}, the differences between the individual
\vsini\ values from each measured line in each spectrum and the
associated mean value for the spectrum are plotted as a function of
$\langle\vsini\rangle$.  
In the same way the error associated with the \vsini\ has been estimated
in Paper~I, a robust estimate of the standard deviation is computed
for each bin of 70 points. The resulting points (open grey circles in
Fig.~\ref{sigma-vsini1}) are adjusted with a linear least squares fit
(dot-dashed line). It gives: 

\begin{equation}
\label{sigma}
\sigma_{\vsini|\vsini} = 0.048{\scriptstyle\pm 0.010}\,\langle\vsini\rangle + 0.14{\scriptstyle\pm 0.19}.\\
\end{equation} 

%
%
%

This fit is carried out using GaussFit \citep{Jes_98a,Jes_98b}, a
general program for the solution of least squares and robust
estimation problems. The resulting constant of the linear fit has an
error bar of the same order than the value itself, and then the formal
error is estimated to be 5~\% of the \vsini.  

\begin{figure}[!htp]
        \centering
\resizebox{\hsize}{!}{\includegraphics{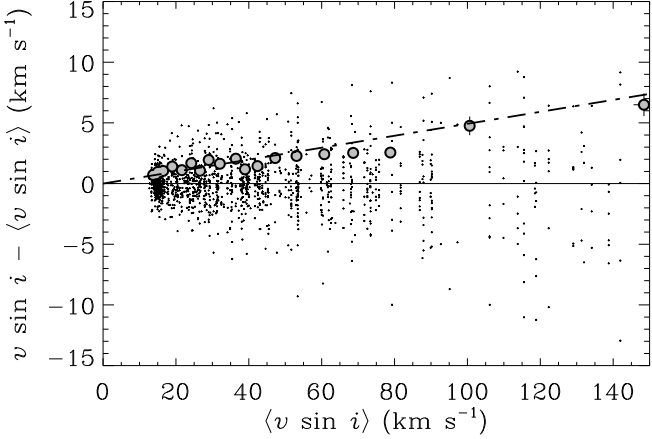}}  
        \caption{Differences between individual \vsini\ and mean over a spectrum $\langle\vsini\rangle$. Variation of the standard deviation associated with the measure with the $\langle\vsini\rangle$ is shown by the open circles. A linear least-square fit on these points (dot-dashed line) gives a slope of 0.05.}
        \label{sigma-vsini1}
\end{figure}

The slope is lower with AUR\'ELIE data than with ECHELEC
spectra (Paper~I): $4.8{\scriptstyle\pm 1.0}$~\% against
$5.9{\scriptstyle\pm 0.3}$~\%. This trend can be explained by the average number of lines for the computation of
the mean \vsini. In the velocity range from 15 to 180~\kms, the number 
of measured lines (Fig.~\ref{line-vsini}) is on average 2.4 times larger with AUR\'ELIE than
with ECHELEC, which could lower the measured dispersion by a factor of $\sqrt{2.4}\approx 1.5$.

\subsubsection{Effect of spectral range} 

As shown in Fig.\ref{histST}, the distribution of spectral types is mainly 
concentrated towards late-B and early-A stars, so that a variation of
the precision as a function of the spectral type would not be very
significant.
On the other hand, as the observed spectral domain is not always the
same, this could introduce an effect due to the different sets of
selected lines, their quantity and their quality in terms of \vsini\
determination. For each of the three spectral domains, the residuals,
normalized by $\sigma_{\vsini|\vsini}$ (Eq.~\ref{sigma}), are centered
around 0 with a dispersion of about 1 taking into account their error
bars, as shown in Table~\ref{sigmaD}. This suggests that no effect due 
to the measurement in one given spectral range is produced on the
derived \vsini.
\begin{table}[!hbtp]
\centering
\caption{Mean of differences between individual \ensuremath{v\sin i}\ and
          average $\langle\ensuremath{v\sin i}\rangle$ over a spectrum, normalized
          by the formal error due to \ensuremath{v\sin i}, are indicated for each
          spectral range as well as the standard deviations $\hat{\sigma}_{\vsini|\Lambda}$
          of these means.}
\label{sigmaD}
\begin{tabular}{ccc}
\hline
Spectral range &
\multicolumn{1}{c}{$\left\langle{\vsini-\langle\ensuremath{v\sin
        i}\rangle\over \sigma_{\vsini|\vsini}}\right\rangle$} & \multicolumn{1}{c}{$\hat{\sigma}_{\vsini|\Lambda}$} \\
\hline
$\Lambda_1$ & $-0.04 {\scriptstyle\pm 0.08}$ & $1.00 {\scriptstyle\pm 0.09}$   \\
$\Lambda_2$ & $-0.10 {\scriptstyle\pm 0.12}$ & $0.83 {\scriptstyle\pm 0.13}$  \\
$\Lambda_3$ & $-0.03 {\scriptstyle\pm 0.03}$ & $0.92 {\scriptstyle\pm 0.04}$  \\
\hline
\end{tabular}
\end{table}
\section{Rotational velocities data} 

\subsection{Results}
\begin{table}[!hbtp]
\centering
\caption{{\bf (extract)} Results of the \vsini\ measurements. Only the 15 first 
stars are listed below. The whole table is available
electronically. Description of the columns is detailed in the text.}
\label{results}
\begin{tabular}{rrlrrrl}
\hline
\multicolumn{1}{c}{HD} & \multicolumn{1}{c}{HIP} & Spect. type & \vsini\ & $\sigma$ & \# & Remark\\
                       &                         &             &  \multicolumn{2}{r}{(\kms)~}  \\
\hline

   905 &   1086 & F0IV        &  35 &  1 &  6 &  \\
  2421 &   2225 & A2Vs        &  14 &  1 &  9 &  \\
  2628 &   2355 & A7III       &  21 &  2 &  9 &  \\
  2924 &   2565 & A2IV        &  31 &  2 & 16 &  \\
  3038 &   2707 & B9III       & 184 & -- &  1 &  \\
  4161 &   3572 & A2IV        &  29 &  2 &  9 &  \\
  4222 &   3544 & A2Vs        &  38 &  2 & 17 &  \\
  4321 &   3611 & A2III       &  25: &  4 & 14 & SS \\
  5066 &   4129 & A2V         & 121 & -- &  1 &  \\
  5550 &   4572 & A0III       &  16 &  3 &  5 &  \\
  6960 &   5566 & B9.5V       &  33 &  4 &  7 &  \\
 10293 &   7963 & B8III       &  62 & -- &  1 &  \\
 10982 &   8387 & B9.5V       &  33 &  3 &  3 &  \\
 11529 &   9009 & B8III       &  36 &  4 &  8 &  \\
 11636 &   8903 & A5V...      &  73 &  2 & 11 &  \\

\hline
\end{tabular}
\end{table}
In total, projected rotational velocities were derived for 249 B8 to
F2-type stars, 86 of which have no rotational velocities in
\citet{AbtMol95}.

The results of the \vsini\ determinations are presented in
Table~\ref{results} which contains the following data: column (1)
gives the HD number, column (2) gives the HIP number, column (3)
displays the spectral type as given in the HIPPARCOS catalogue
\citep{Hip}, columns (4, 5, 6) give respectively the derived value of
\vsini, the associated standard deviation and the corresponding number
of measured lines (uncertain \vsini\ are indicated by a colon), column
(7) presents possible remarks about the spectra: SB2 (``SB'') and
shell (``SH'') natures are indicated for stars showing such feature in 
these observed spectra, as well as the reason why \vsini\ is uncertain -- ``NO''
for no selected lines, ``SS'' for variation from spectrum to spectrum
and ``LL'' for variation from line to line (see Appendix~\ref{Notes0}).

\subsubsection{SB2 systems} 
Nine stars are seen as double-lined spectroscopic binary in the data
sample. Depending on the \vsini\ of each component, their difference
in Doppler shift and their flux ratio, determination of \vsini\ is
impossible in some cases.

\begin{figure*}[!htp]
        \centering
\resizebox{\hsize}{!}{\includegraphics{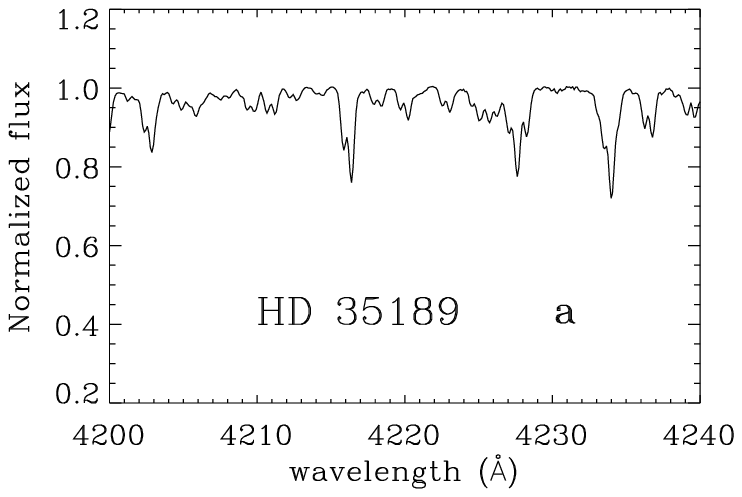}\includegraphics{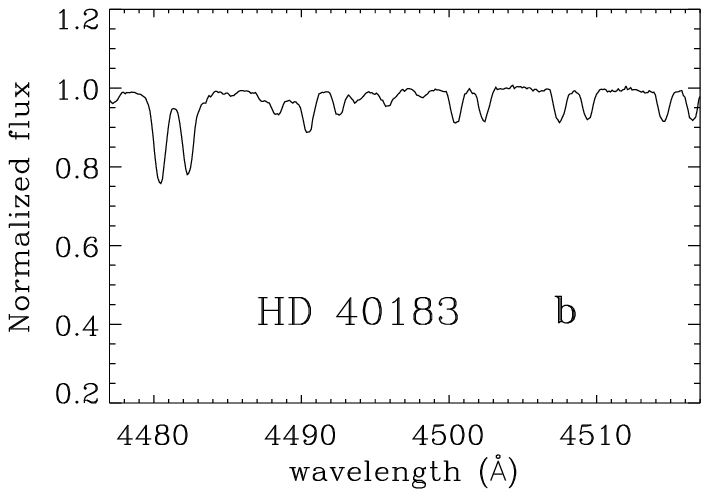}\includegraphics{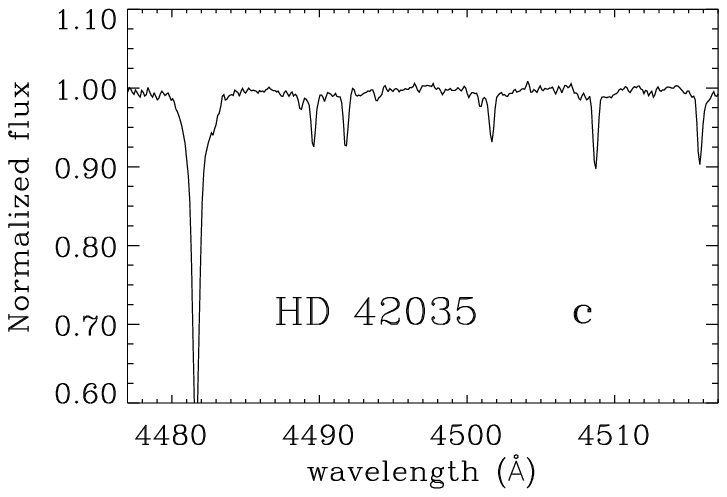}}
\resizebox{\hsize}{!}{\includegraphics{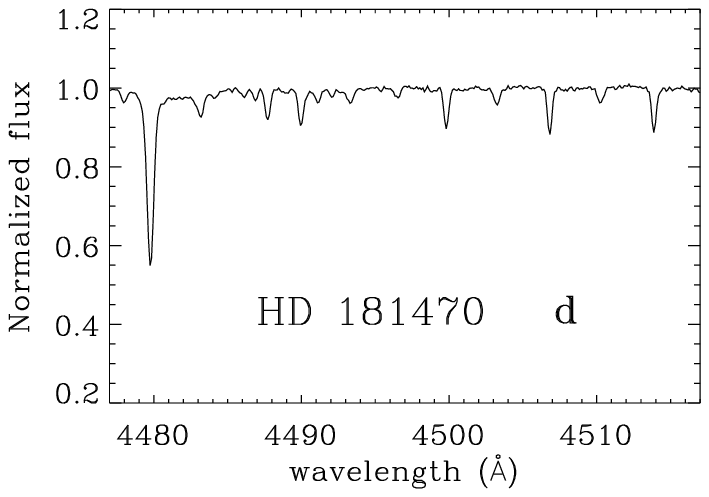}\includegraphics{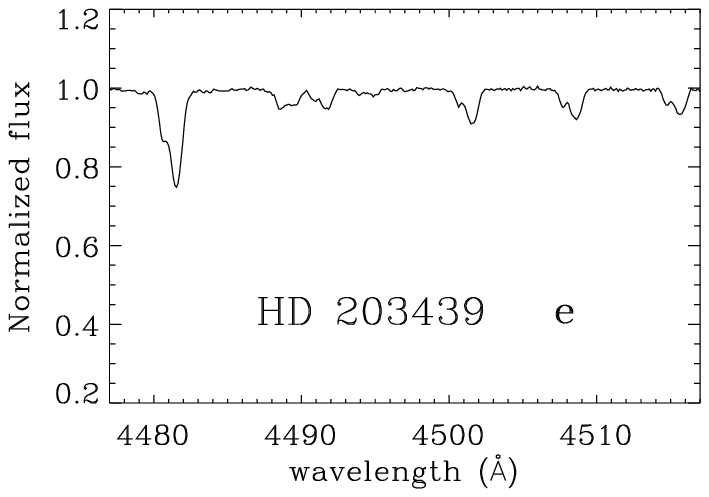}\includegraphics{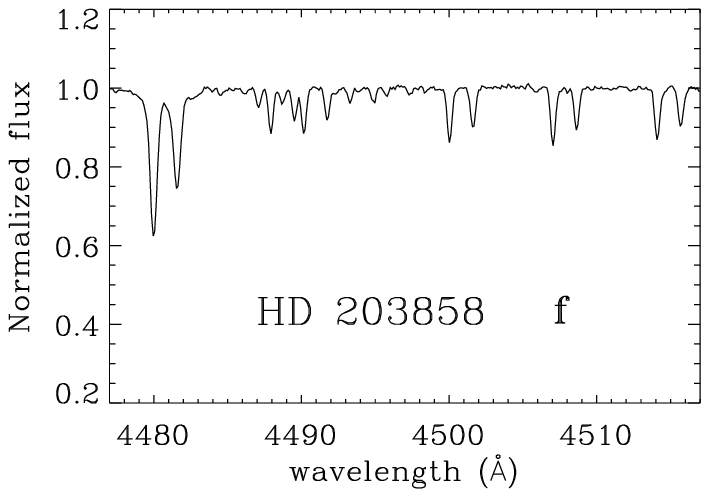}}  
\caption{Part of the spectra are displayed for the six SB2 stars that
  have been observed only once: {\bf a)} HD~35189, {\bf b)} HD~40183,
  {\bf c)} HD~42035, {\bf d)} HD~181470, {\bf e)} HD~203439, {\bf f)}
  HD~203858. Three of them are well separated (b, d, f), allowing
  measurement of \vsini\ for both components. The three others (a, c,
  e) have low differential Doppler shift ($\le 60$~\kms) which makes
  all the lines blended. No \vsini\ has been determined for these objects.}

        \label{sb2_7}
\end{figure*}


\begin{figure*}[!htp]
        \centering
\resizebox{\hsize}{!}{\includegraphics{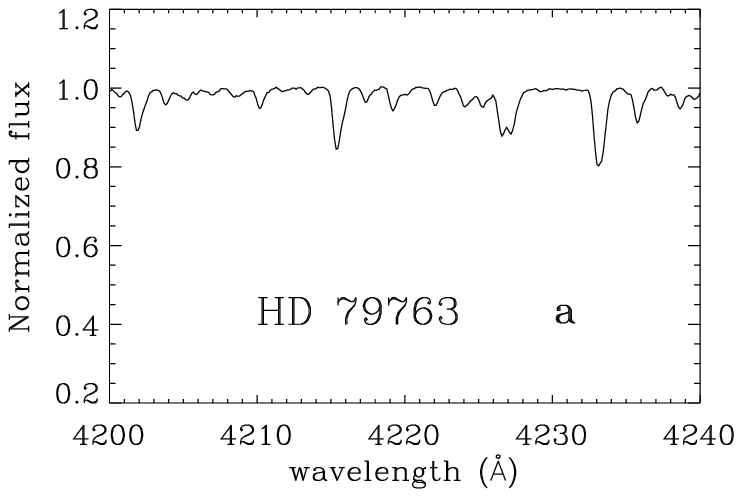}\includegraphics{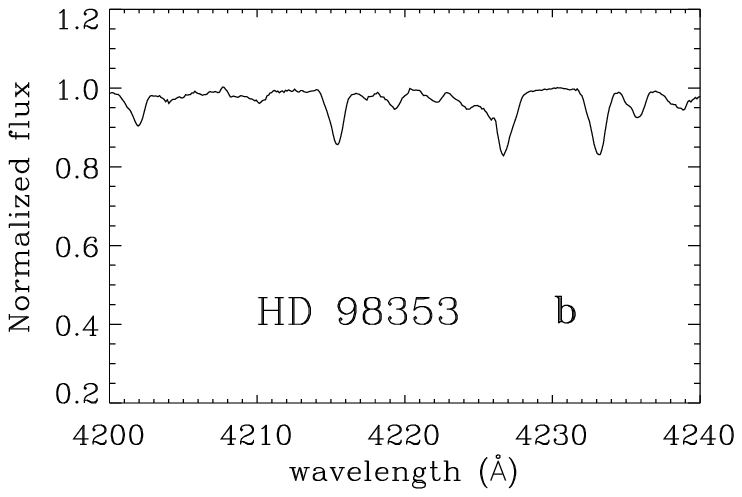}\includegraphics{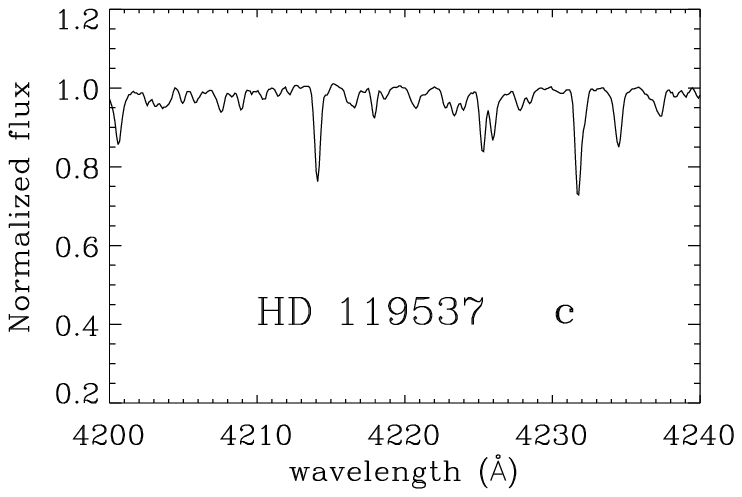}}
\resizebox{\hsize}{!}{\includegraphics{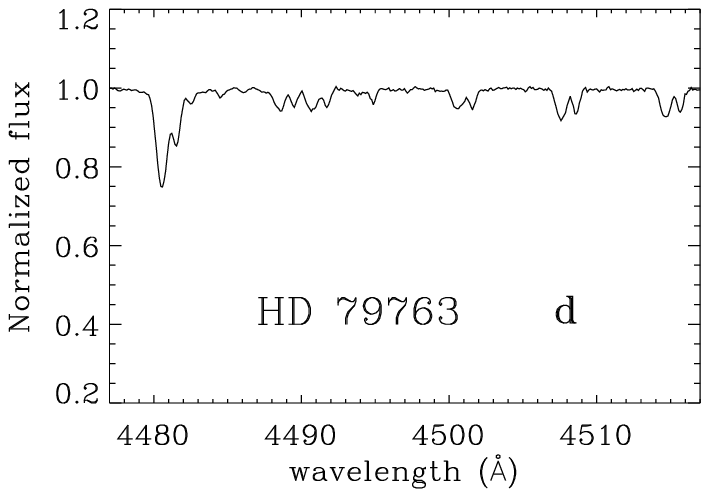}\includegraphics{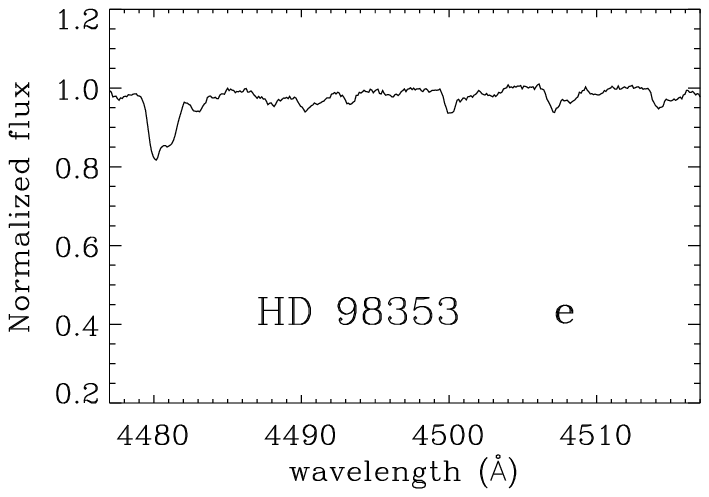}\includegraphics{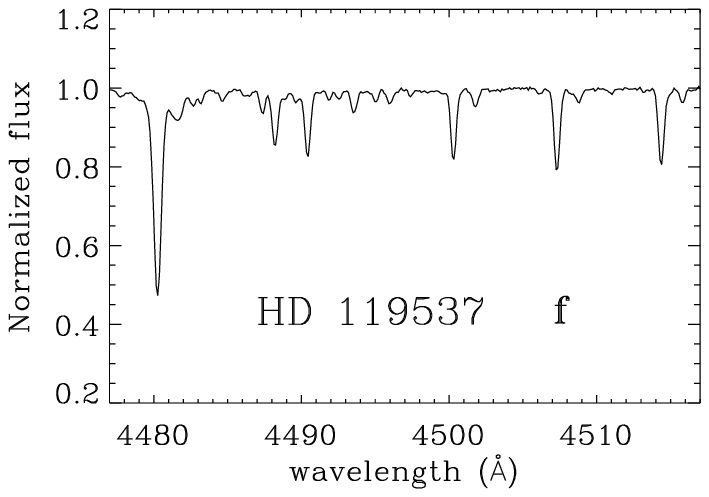}}  
 \caption{The three following SB2 stars have been observed twice, in
   $\Lambda_1$ (upper panels) and $\Lambda_3$ (lower panels): {\bf a)}
   HD~79763 at HJD 2449025, {\bf b)} HD~98353 at HJD 2448274,  {\bf c)}   HD~119537 at HJD 2449025, {\bf d)}
   HD~79763 at HJD 2449365, {\bf e)} HD~98353 at HJD 2449413,  {\bf f)}
   HD~119537 at HJD 2449415. SB2 nature of these objects is not
   detected in $\Lambda_1$ spectral range, and the derived \vsini\ is
   a ``combined'' broadening. The triple system HD~98353 is observed
   close to conjunction, and lines remain blended. For HD~79763 (d)
   and  HD~119537 (f), the difference in radial velocity is large
   enough to measure separately the rotational velocities.}
%
\label{sb2_5}
\end{figure*}



Table~\ref{SB2} displays the results for the stars in our sample which
exhibit an SB2 nature. Spectral lines are identified by
comparing the SB2 spectrum with a single star spectrum. Projected
rotational velocities are given for each component when measurable, as
well as the difference in radial velocity $\Delta V_\mathrm{r}$
computed from a few lines in the spectrum.
\begin{table}[!htp]
\centering
\caption{Results for stars seen as SB2. Rotational velocities are
  given for each component when measurable. $\Delta V_\mathrm{r}$
  stands for the difference in radial velocity between the two
  components. Dash indicates a non
  possible measurement (either for \vsini\ or $\Delta V_\mathrm{r}$).}
\label{SB2}
\setlength\tabcolsep{5pt}
\begin{tabular}{rrlrrrc}
\hline
\multicolumn{1}{c}{HD}  & \multicolumn{1}{c}{HIP} & Spect. type & \multicolumn{2}{c}{\vsini} & $\Delta V_\mathrm{r}$ & Fig.  \\
    &    &    & \multicolumn{2}{c}{(\kms)}&\multicolumn{1}{c}{(\kms)}  \\
    &    &    & \multicolumn{1}{c}{A}&\multicolumn{1}{c}{B}  \\
\hline
 35189 &  25216 & A2IV      & \multicolumn{2}{c}{--}  &  37 & \ref{sb2_7}a. \\
 40183 &  28360 & A2V       &   37 &  37              & 127 & \ref{sb2_7}b. \\
 42035 &  29138 & B9V       & \multicolumn{2}{c}{see text}           &  12:& \ref{sb2_7}c.\\
 79763 &  45590 & A1V       & \multicolumn{2}{c}{29}  & --  & \ref{sb2_5}a.\\
       &        &           &   34: &  21:              &  67 & \ref{sb2_5}d.\\
 98353 &  55266 & A2V       & \multicolumn{2}{c}{44}  &   & \ref{sb2_5}b.\\
       &        &           & \multicolumn{2}{c}{34}  & 64: & \ref{sb2_5}e.\\
119537 &  67004 & A1V       & \multicolumn{2}{c}{20:}  & --  & \ref{sb2_5}c. \\
       &        &           &   17 &  18  &  98 & \ref{sb2_5}f. \\
181470 &  94932 & A0III     &   15 &  20  & 229 & \ref{sb2_7}d. \\
203439 & 105432 & A1V       &  \multicolumn{2}{c}{--} & 56  & \ref{sb2_7}e. \\
203858 & 105660 & A2V       &   14 &  15  & 106 & \ref{sb2_7}f. \\
\hline
\end{tabular}
\end{table}

\begin{itemize}
\item \object{110~Tau} (HD~35189) is highly suspected to be a
  spectroscopic binary due to its spectra taken in the 
  $\Lambda_1$ domain. As it can be seen in Fig.~\ref{sb2_7}.a, core of the
  lines is double in most cases, corresponding to a difference in
  radial velocity of about 37~\kms. However, this case does not allow the measurement of the projected rotational velocity.
\item \object{$\beta$~Aur} (HD~40183) is a well known early A-type eclipsing binary 
  of Algol type. The derived \vsini\ corresponds well with the values
  of \citet{NomJon94} (respectively 33 and 34~\kms) that indicate
  synchronous rotation. Knowing the time of minimum from \citet{JonSon97}:
 $$\mathrm{HJD}_\mathrm{min} = 2431076.7269 + 3.96004732\, E, $$ 
 where $E$ is an integer of phases,
 and the Julian date of observation (HJD 2449413.3301), the phase is equal to 0.4.
\item \object{HD~42035} has no indication of any binary status in the
  literature and it is flagged as a photometrically constant star from
  HIPPARCOS data. Nevertheless the observed line profiles in its
  spectrum tend to suggest that it is composite. The cross-correlation function (CCF) of the observed spectrum with a synthetic one
  ($\Teff = 10\,000$~K, $\logg =4.0$, $\vsini=5$~\kms) has been computed. This CCF is not 
  characteristic of a single star, and it is perfectly fitted by the
  sum of two gaussian components centered at 26 and 34~\kms\
  respectively and whose $FWHM$ are 30 and 120~\kms.
This indicates that the system is composed of a low \vsini\ star
and a faster rotator. 
\item \object{HD~79763} is known as a SB2 system whose orbital period
  is $P=15.986$~d \citep{Ban_89}. Using the spectrum in $\Lambda_3$
  range (Fig.~\ref{sb2_5}.d), the difference in radial velocity is sufficient to estimate
  \vsini\ of both components.
\item \object{55~UMa} (HD~98353) is a triple system for which components
  are early A-type stars. Using tomographic separation, \citet{Liu_97}
  estimate the \vsini\ of each of them: $30\pm 4$~\kms, $45\pm
  5$~\kms, $55\pm 5$~\kms. Knowing the orbital parameters:
  $P=2.5538380$~d and $T_\mathrm{minRV} = 2449602.588$
  \citep{Hon_96} for the close pair, our spectra correspond to phases
  $\phi=0.99$ (Fig.~\ref{sb2_5}.b, HJD 2448274.6233) and
  $\phi=0.02$ (Fig.~\ref{sb2_5}.e, HJD 2449413.5500). This means that 
   both observations were unfortunately made close to opposition, and
  the difference in radial velocity is not large enough to see separated
  lines. The measured \vsini\ corresponds to a blend.
\item \object{HD~119537} is indicated as SB in the Bright Star Catalogue
  \citep{Bsc1} and was not detected as a double-lined system with the spectrum collected 
  at ESO, within the context of the southern sample \citep{Grr_99a,
    Ror_02a}. The rotational velocity derived in Paper~I is $13\pm
  1$~\kms, whereas it equals 23.9~\kms\ in \citet{Raa_89}. The spectrum in the $\Lambda_3$ domain displays evidence of 
  SB2 nature and Fig.~\ref{sb2_5}.f shows perfectly the faint lines
  around \ion{Ti}{ii}~4501, \ion{Fe}{ii}~4508 and
  \ion{Ti}{ii}~4515 which are usually well isolated in a single star
  with such a spectral type.

\item \object{HD~181470} is a close binary system first detected by
  speckle observations by \citet{Mia_93} and later confirmed by
  \citet{Haf_00}. The measured separation is $\rho= 0\farcs 13$ with a 
  magnitude difference  $\Delta m= 1.6\pm 0.2$. In the observed
  spectrum, the difference in radial velocity is large: $229$~\kms.

\item \object{HD~203439} is known as a spectroscopic binary system in \citet{Ban_89}.

\item \object{HD~203858} is a known spectroscopic
  binary. \citet{AbtMol95} do not see the two components of the system
  and give a \vsini\ which is likely overestimated, 70~\kms, because of blend
  due to binarity. In this case, the components are well separated and 
  each \vsini\ is measured.
\end{itemize}

\subsection{Comparison with existing data} 

\subsubsection{South versus North} 
\label{so-no}
Fourteen stars are common to both the southern sample from Paper~I
and the northern one studied here. Matching of both determinations allows
us to ensure the homogeneity of the data or indicate variations intrinsic 
to the stars otherwise. Results for these objects are listed in
Table~\ref{vsini_NS}.
\begin{table}[!h]
\begin{center}
\caption{Comparison of the computed \vsini\ for the stars in common in 
  the northern and southern samples (N $\equiv$ this work, S $\equiv$ Paper~I). CFF
  is a flag indicating the shape of the cross-correlation function
  carried out by \citet{Grr_99a} using the ECHELEC spectra (0: symmetric and gaussian peak, 4: probable double, 5: suspected double, 6: probable multiple system).}
\label{vsini_NS}
\begin{tabular}{rlcrlrl}
\hline
  HD   & Sp. type & CCF & $\vsini_{\mathrm{N}}$& $\sigma_{\mathrm{N}}$& $\vsini_{\mathrm{S}}$& $\sigma_{\mathrm{S}}$ \cr
\hline

 27962 & A2IV     & 0   & 16  & 2  &  11  & 1   \cr
 30321 & A2V      & 4   &132  & 4  & 124  & --  \cr
 33111 & A3IIIvar & 6   &196  & -- & 193  & 4   \cr
 37788 & F0IV     & 0   & 29  & 1  &  33  & 4   \cr
 40446 & A1Vs     & --  & 27  & 5  &  27  & 5   \cr
 65900 & A1V      & 0   & 35  & 3  &  36  & 2   \cr
 71155 & A0V      & 4   &161  &12  & 137  & 2   \cr
 72660 & A1V      & 0   & 14  & 1  &   9  & 1   \cr
 83373 & A1V      & 0   & 28  & -- &  30  & 2   \cr
 97633 & A2V      & 0   & 24  & 3  &  23  & 1   \cr
 98664 & B9.5Vs   & --  & 57  & 1  &  61  & 5   \cr
109860 & A1V      & 5   & 74  & 1  &  76  & 6   \cr
193432 & B9IV     & 0   & 24  & 2  &  25  & 2   \cr
198001 & A1V      & 0   &130  & -- & 102  & --  \cr
\hline
\end{tabular}
\end{center}
\end{table}
 
Instrumental characteristics differ from ECHELEC to AUR\'ELIE
data. First of all, the resolution is higher in the ECHELEC spectra,
which induces a narrower instrumental profile and allows the
determination of \vsini\ down to a lower limit. Taking the calibration
relation from SCBWP as a rule of thumb ($FWHM{\scriptstyle
  [\mathrm{\AA}]} \approx 0.025\,\vsini{\scriptstyle [\kms]}$), the
low limit of \vsini\ is:
\begin{equation}
\label{vsinilim}
\vsini_\mathrm{lim} = {1\over 0.025}\, FWHM_\mathrm{inst} = {1\over 0.025}\, {\lambda\over R},
\end{equation} 
where $R$ is the power of resolution, and $\lambda$ the considered
wavelength. For ECHELEC spectra ($R\approx 28\,000$) this limit is
6.4~\kms\ at 4500~\AA, whereas for AUR\'ELIE data ($R\approx 16\,000$), it reaches
11.3~\kms. These limits correspond to the ``rotational velocity'' associated with the $FWHM$
of the instrumental profile. There is no doubt that in Fourier space, the
position of the first zero of a line profile dominated by the
instrumental profile is rather misleading and the effective lowest
measurable \vsini\ may be larger. This effect explains the discrepancy found for slow
rotators, i.e. HD~27962 and HD~72660 in Table~\ref{vsini_NS}. 
The \vsini\ determination using AUR\'ELIE spectra is 5~\kms\ larger
than using ECHELEC spectra.
The two stars are slow rotators for which \vsini\ has already been
derived using better resolution. HD~27962 is found to have $\vsini=12$ 
and $11$~\kms\ by \citet{VaeMor99} and \citet{HuaAln98}
respectively. HD~72660 has a much smaller \vsini, lower than the limit 
due to the resolution of our spectra: 6.5~\kms\ in \citet{NinWan00}
and 6~\kms\ in \citet{Vae99}.

Second of all, one other difference lies in the observed spectral
domain. HD~198001 has no observation in the $\Lambda_3$ domain using
AUR\'ELIE, so that $\vsini_\mathrm{N}$ in Table~\ref{vsini_NS} is not derived
on the basis of the \ion{Mg}{ii} line. The overestimation of
$\vsini_\mathrm{N}$ reflects the use of weak metallic lines instead the strong
\ion{Mg}{ii} line for determining rotational velocity. 

Using the same ECHELEC data, \citet{Grr_99a} flagged the stars according to
the shape of their cross-correlation function with synthetic
templates. This gives a hint about binary status of the stars. Three
stars in Table~\ref{vsini_NS} are flagged as ``probable binary or
multiple systems'' (CCF: 4 and 6).

When discarding low rotators, probable binaries and data of HD~198001 that induce biases
in the comparison, the relation between the eight remaining points is fitted
using GaussFit by:
\begin{equation}
\label{N-S}
\vsini_\mathrm{S} = 1.05{\scriptstyle\pm 0.04}\,\vsini_\mathrm{N}-0.2{\scriptstyle\pm 1.5}.
\end{equation} 

Although common data are very scarce, they seem to be consistent. It
suggests that both data sets can be merged as long as great care is
taken for cases detailed above, i.e. extremely low rotators, high
rotators with no \vsini\ from \ion{Mg}{ii} line, spectroscopic binaries.

\subsubsection{Standard stars} 
\label{comparison}

\begin{figure}[!htp]
\resizebox{\hsize}{!}{\includegraphics{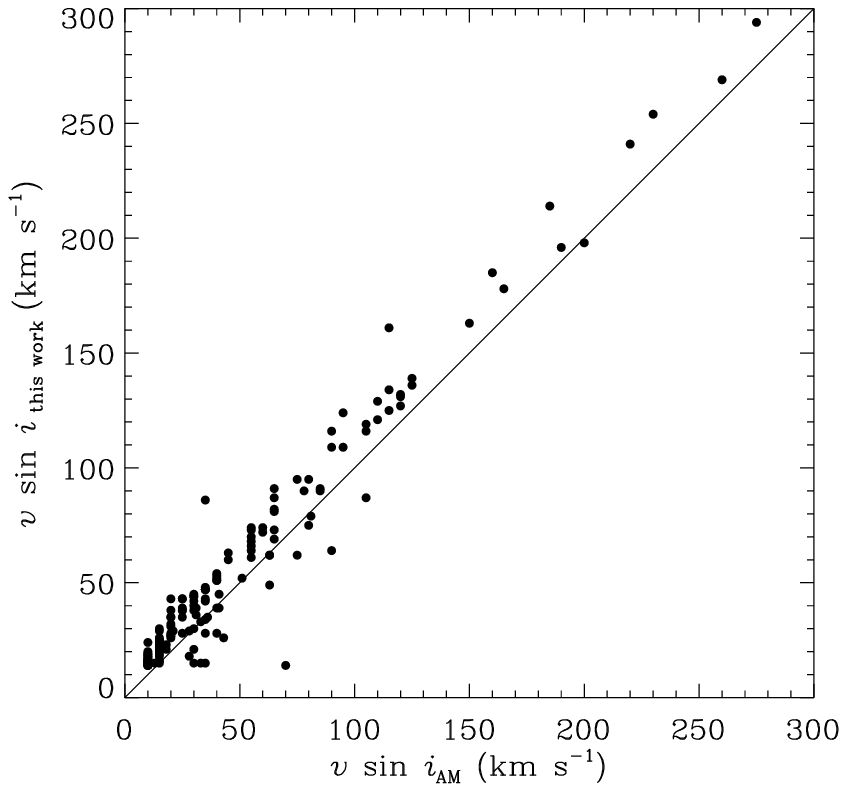}}
\hfill
\caption{Comparison of \vsini\ data for the 163 common stars between this work and 
  \citet{AbtMol95}. The solid line stands for the one-to-one
  relation. }
\label{comp-vsini1}
\end{figure}

\begin{figure}[!htp]
\resizebox{\hsize}{!}{\includegraphics{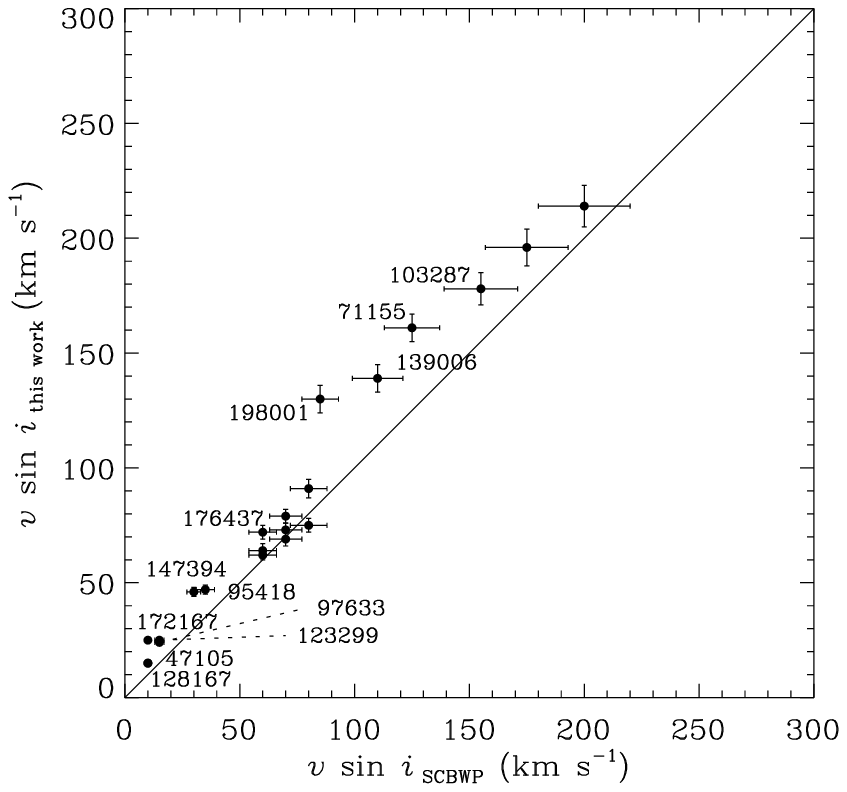}}
\hfill
\caption{Comparison between \vsini\ data from this work and from
  \citet{Slk_75}. The solid line stands for the one-to-one
  relation. The 21 standard stars are plotted with error bar on both
  axes (see text). HD number of the stars that deviate most from the one-to-one
  relation are indicated and these stars are listed in Table~\ref{vsini_litt} and detailed in Appendix~\ref{Notes3}.}
\label{comp-vsini}
\end{figure}
A significant part of the sample is included in the catalogue of
\cite{AbtMol95}. The intersection includes 163 stars. The
comparison of the \vsini\ (Fig.~\ref{comp-vsini1}) shows that our determination is higher on average than
the velocities derived by Abt \& Morrell (AM). The linear relation given by GaussFit is:
\begin{equation}
\label{Ram-AM}
\vsini_\mathrm{this\;work} = 1.18{\scriptstyle\pm 0.04}\,\vsini_\mathrm{AM}+3.8{\scriptstyle\pm 0.8}.
\end{equation}
Abt \& Morrell use the standard stars of SCBWP to calibrate the relation $FWHM$--\vsini. There are 21 stars in common between our sample and these standard stars. 
Figure~\ref{comp-vsini} displays the \vsini\ derived in this paper
versus the \vsini\ from SCBWP for these  21 common stars. The solid line represents the one-to-one relation. A clear trend is observed: \vsini\ from SCBWP are on average 20~\% lower. A linear least squares fit carried out with GaussFit on these values makes the systematic effect explicit:

\begin{equation}
\label{Ram-Slk}
\vsini_\mathrm{this\;work} = 1.11{\scriptstyle\pm 0.07}\,\vsini_\mathrm{SCBWP}+7.1{\scriptstyle\pm 1.7}.
\end{equation}

The relation is computed taking into account the error bars of both sources. The error bars on the values of SCBWP are assigned according to the accuracy given in their paper (10\,\% for $\vsini<200\,\kms$ and 15\,\% for $\vsini\geq 200\,\kms$). Our error bars are derived from the formal error found in section~\ref{precision} (Eq.~\ref{sigma}).

  \begin{table*}
\begin{center}
\caption{Highlight of the discrepancy between \vsini\ values from SCBWP and ours (standard deviation of our measurement is indicated; dash ``--'' stands for only one measurement). Comparison with data from the literature for the twelve stars that exhibit the largest differences. \vsini\ are classified in three subgroups according to the way they are derived: by-product of a spectrum synthesis, frequency analysis of the lines profiles or infered from a $FWHM$-\vsini\ relation independent from SCBWP's one. Flags from HIPPARCOS catalogue are indicated: variability flag H52 (C: constant, D: duplicity-induced variability, M: possibly micro-variable, U: unsolved variable, --: no certain classification) and double annex flag H59 (O: orbital solution, G: acceleration terms, --: no entry in the Double and Multiple Systems Annex).}
\label{vsini_litt}
\begin{tabular}{lrlrr@{\hspace*{1mm}}llllcc}
\hline
\multicolumn{1}{c}{Name} & \multicolumn{1}{c}{HD} &
\multicolumn{1}{c}{Sp. type} &  \multicolumn{6}{c}{\vsini\ (\kms)}
& \multicolumn{2}{c}{{\sc hipparcos}}\cr
   &    &      &  {\sc scbwp} & \multicolumn{2}{c}{this
     work}&\multicolumn{3}{c}{\hbox{\raisebox{0.3em}{\vrule depth 0pt
         height 0.4pt width 3.0cm} literature \raisebox{0.3em}{\vrule
         depth 0pt height 0.4pt width 3.0cm}}} & H52 & H59\cr
   &     &      &    &    &      & \multicolumn{1}{c}{spec. synth.}& \multicolumn{1}{c}{freq. analysis}& \multicolumn{1}{c}{$FWHM$}\cr
\hline
$\gamma$~Gem   & 47105& A0IV    &  $<10$ & $ 15$ & ${\scriptstyle\pm
  1}$ &$11.2^{(1)}$        & $10.2{\scriptstyle\pm 0.2}^{(2)}$, $19.0^{(3)}$    &      & -- & X  \cr
30~Mon         & 71155& A0V     &  $125$ & $161$ & ${\scriptstyle\pm 12}$ &                    &                                      &      & C  & -- \cr
$\beta$~UMa    & 95418& A1V     &   $35$ & $ 47$ & ${\scriptstyle\pm  3}$ &$44.8^{(1)}$, $39^{(4)}$        &  $44.3^{(3)}$         &      & -- & -- \cr
$\theta$~Leo   & 97633& A2V     &   $15$ & $ 24$ & ${\scriptstyle\pm 3}$ &$21^{(5)}$, $22.1^{(1)}$ & $24^{(6)}$, $27.2^{(3)}$    & $23^{(7)}$       & -- & -- \cr
$\gamma$~UMa   &103287& A0V SB  &  $155$ & $178$ & ${\scriptstyle\pm  9}$ &      & $154{\scriptstyle\pm 4}^{(8)}$     &      & M & --   \cr
$\alpha$~Dra   &123299& A0III SB&   $15$ & $ 25$ & ${\scriptstyle\pm  2}$ &      & $27^{(9)}$     &      & M & O  \cr
$\sigma$~Boo   &128167& F3Vwvar &   $10$ & $ 15$ & ${\scriptstyle\pm 1}$ & $7.5{\scriptstyle\pm 1}^{(10)}$ & $7.5^{(11)}$ &$7.8^{(12)}$, $8.1^{(13)}$     & --  & --   \cr
$\alpha$~CrB   &139006& A0V     &  $110$ & $139$ & ${\scriptstyle\pm 10}$ &      & $127{\scriptstyle\pm 4}^{(8)}$     &      & U & O  \cr
$\tau$~Her     &147394& B5IV    &   $30$ & $ 46$ & ${\scriptstyle\pm  3}$ &      & $32^{(6)}$     &      & P &  --  \cr
$\alpha$~Lyr   &172167& A0Vvar  &  $<10$ & $ 25$ & ${\scriptstyle\pm 2}$ &$22.4^{(1)}$, $23.2^{(14)}$  & $23.4{\scriptstyle\pm 0.4}^{(16)}$, $24^{(6)}$     &      & U &  --  \cr
               &      &         &        &       &             &$21.8{\scriptstyle\pm 0.2}^{(15)}$  & $29.9^{(3)}$ \cr
$\gamma$~Lyr   &176437& B9III   &   $60$ & $ 72$ & ${\scriptstyle\pm  2}$ &      &      &      & M &  --  \cr
$\epsilon$~Aqr &198001& A1V     &   $85$ & $130$ & -- & $95^{(17)}$, $108.1^{(1)}$     &      &      &  -- &  --  \cr

\hline
\end{tabular}
\end{center}
\begin{tabular}{llll}
$^{(1)}$ \citet{Hil95}    &  $^{(6)}$ \citet{SmhDwy93} & $^{(11)}$  \citet{Gry84}  & $^{(16)}$ \citet{Gry80a}\\
$^{(2)}$ \citet{Scz_97}   &  $^{(7)}$ \citet{Fel98}    & $^{(12)}$ \citet{Fel97}   & $^{(17)}$ \citet{Dun_97}\\
$^{(3)}$ \citet{Raa_89}   &  $^{(8)}$ \citet{Gry80b}   & $^{(13)}$ \citet{BezMar84}  \\
$^{(4)}$ \citet{Hor_99}   &  $^{(9)}$ \citet{LenScz93} & $^{(14)}$ \citet{ErrNoh02}  \\
$^{(5)}$ \citet{Lee89}    &  $^{(10)}$ \citet{Som82}   & $^{(15)}$ \citet{Gur_94}    \\
\end{tabular}
\end{table*}

The standard stars for which a significant
discrepancy occurs between our values and those derived by SCBWP
 -- i.e. their error box does not intersect with the one-to-one
  relation -- have their names indicated in Fig.~\ref{comp-vsini}. They are
listed with data from the literature in Table~\ref{vsini_litt} and
further detailed in Appendix~\ref{Notes3}.

\section{Merging the samples}
\label{merge}
Homogeneity and size are two crucial characteristics of a sample, in a 
statistical sense.  In order to gather a \vsini\ sample obeying these
two criteria, \vsini\ derived in this paper and in Paper~I can be
merged with those of \citet{AbtMol95}. The different steps consist of
first joining the new data, taking care of their overlap; then
considering the intersection with Abt \& Morrell, carefully scaling
their data to the new ones; and finally gathering the complete homogenized sample.

\subsection{Union of data sets from Paper~I and this work (I $\cup$ II)}
Despite little differences in the observed data and the way \vsini\
were derived for the two samples, they are consistent. The gathering
contains 760 stars. Rotational velocity of common stars listed in
Table~\ref{vsini_NS} are computed as the mean of both values, weighted 
by the inverse of their variance. This weighting is carried on when
both variances are available (i.e. $\sigma_{\mathrm{N}}^2$ and
$\sigma_{\mathrm{S}}^2$), except for low rotators and HD~198001, for
which $\vsini_{\mathrm{S}}$ is taken as the retained value.

\subsection{Intersection with Abt \& Morrell and scaling} 
In order to adjust by the most proper way the scale from Abt \& Morrell's data to the one
defined by this work and the Paper~I, only non biased \vsini\ should
be used. The common subsample has to be cleaned from spurious
determinations that are induced by the presence of spectroscopic binaries,
the limitation due to the resolution, uncertain velocities of high rotators with no
measurement of the \ion{Mg}{ii} doublet, etc.
The intersection gathers 308 stars, and Fig.~\ref{comp-vsini3} displays the comparison. 

We have chosen to adjust the scaling from Abt \& Morrell's data (AM) to
ours (I $\cup$ II) using an iterative linear regression with sigma clipping.
The least-squares linear fit is computed on the data, and the
relative difference 
$$\Delta = \left(\vsini_{\mathrm{I}\cup
    \mathrm{II}}-(A\,\vsini_{\mathrm{AM}}+B)\right)/\vsini_{\mathrm{I}\cup \mathrm{II}},$$ where $A$ and $B$ are the coefficients of the regression line, 
is computed for each point. The standard deviation $\sigma_\Delta$ of all these
differences is used to reject aberrant points, using the criterion :
$$|\Delta| > 1.1\,\sigma_\Delta.$$ Then, the least-squares linear
fit is computed on retained points and the sigma-clipping is repeated until no new points are rejected.  
One can see in previous section that points lying one sigma beyond
their expected value are already significantly discrepant, this
reinforces the choice of the threshold $1.1\,\sigma_\Delta$.

The 23 points rejected during the sigma-clipping iterations are
indicated in Fig.~\ref{comp-vsini3} by open symbols. They are listed
and detailed in Appendix~\ref{Notes4}. Some of them are known as
spectroscopic binaries.
Moreover, using HIPPARCOS
data, nine of the rejected stars are indicated as ``duplicity induced
variable'', micro-variable or double star.
Half a dozen stars are low \vsini\ stars observed with AUR\'ELIE, and
the resolution limitation can be the source of the discrepancy

\begin{figure*}[!htp]
\resizebox{12cm}{!}{\includegraphics{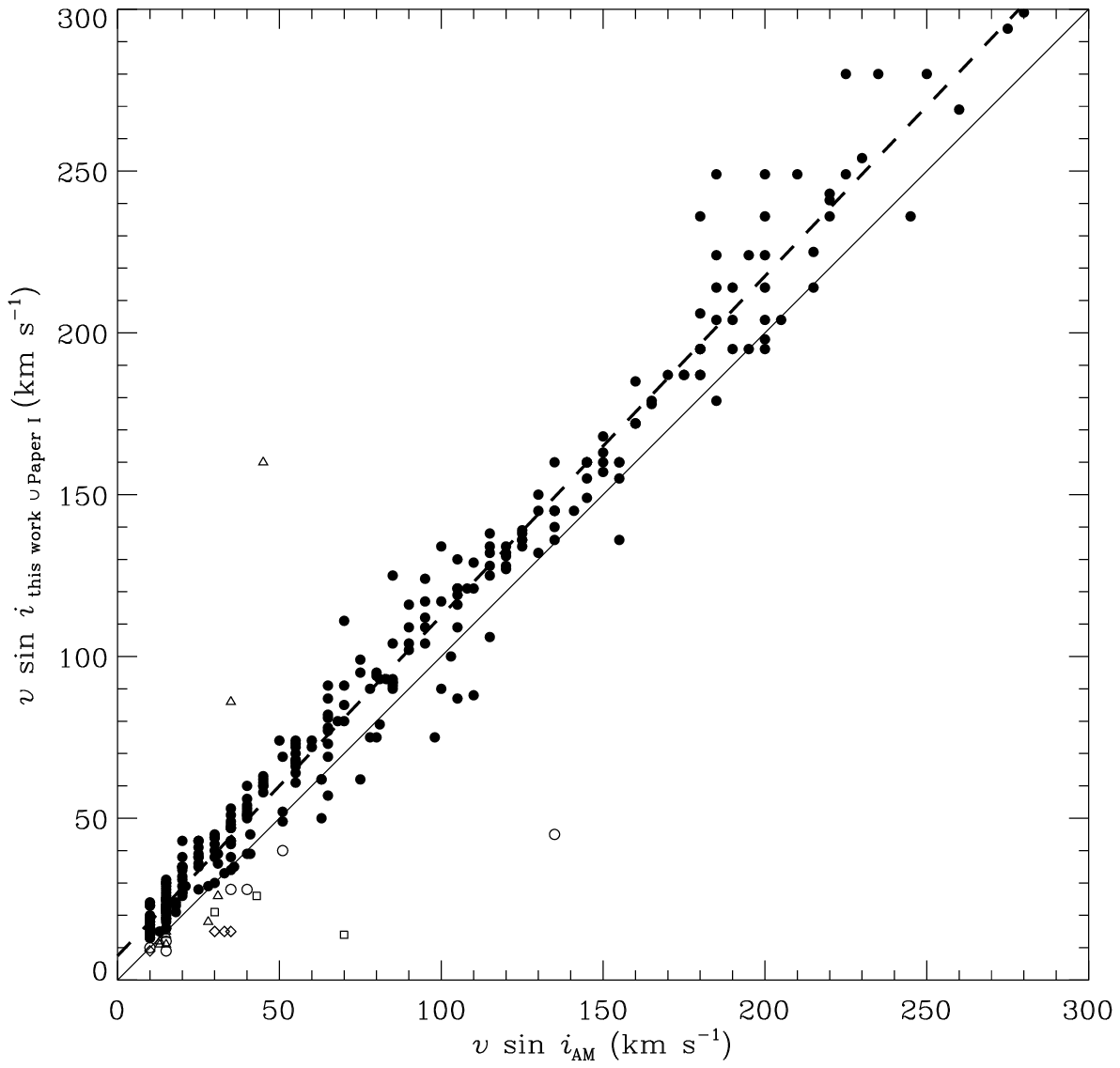}}
\hfill
\parbox[b]{55mm}{  
\caption{Comparison of \vsini\ data for the 308 common stars between 
  \citet{AbtMol95} and the union of data from this work and from Paper~I. 
  Filled circles stand for stars from the ``cleaned'' intersection,
  that are used in the fit of Eq.~\ref{scale},
whereas open symbols represent stars discarded from the scaling fit
(see text). 
The different open symbols indicate the possible reason why the
corresponding stars are discarded: open square: known spectral binary
system; open triangle:
variability flag (H52) or binary flag (H59) in HIPPARCOS; open diamond: very low \vsini\
from AUR\'ELIE data; open circle: no reason.
 The solid line stands for the one-to-one
  relation.  The dashed line is the fit carried on filled circles.
All the discarded objects (open symbols) are listed and detailed in Appendix~\ref{Notes4}.}
\label{comp-vsini3}
}
\end{figure*}
The ``cleaned'' intersection, gathering 285 stars, is represented in
Fig.~\ref{comp-vsini3} by filled circles. The solid line is the one-to-one relation and
the dashed line represents the relation given by the iterative linear fit:
\begin{equation}
\label{scale}
\vsini_{\mathrm{I}\cup \mathrm{II}} = 1.05\,\vsini_\mathrm{AM}+7.5.
\end{equation} 

Rotational velocities from Abt \& Morrell are scaled to the \vsini\
derived by Fourier transform (union of data sets from Paper~I and this work), according to
Eq.~\ref{scale}, in order to merge homogeneous data.

\subsection{Final merging} 

Table~\ref{merging} lists the 2151 stars in the total merged sample. 
It contains the following data: column (1)
gives the HD number, column (2) gives the HIP number, column (3)
displays the spectral type as given in the HIPPARCOS catalogue
\citep{Hip}, column (4) gives the derived value of
\vsini\ (uncertain \vsini, due to uncertain determination in either
one of the source lists, are indicated by a colon).
\begin{table}[!htp]
\centering
\caption{{\bf (extract)} Results of the merging of \vsini\ samples. Only the 15 first 
stars are listed below. The whole table is available
electronically. $\in$ stands for the membership and flags which sample stars belong to: 1,
sample from Paper~I; 2, sample from this work; 4, sample from
\citet{AbtMol95}. This flag is set bitwise, so multiple membership is
set by adding values together.}
\label{merging}
\begin{tabular}{rrlrc}
\hline
\multicolumn{1}{c}{HD} & \multicolumn{1}{c}{HIP} & Spect. type & \vsini & $\in$ \\
                       &                         &             &  (\kms)\\
\hline
    3 &  424 & A1Vn	  & 228 & 4  \\
  203 &  560 & F2IV	  & 170 & 4  \\
  256 &  602 & A2IV/V	  & 241 & 5  \\
  315 &  635 & B8IIIsp... &  81 & 4  \\
  319 &  636 & A1V	  &  59 & 5  \\
  431 &  760 & A7IV	  &  97 & 4  \\
  560 &  813 & B9V	  & 249 & 1  \\
  565 &  798 & A6V	  & 149 & 1  \\
  905 & 1086 & F0IV	  &  36 & 6  \\
  952 & 1123 & A1V	  &  75 & 4  \\
 1048 & 1193 & A1p	  &  28 & 4  \\
 1064 & 1191 & B9V	  & 128 & 1  \\
 1083 & 1215 & A1Vn	  & 233 & 4  \\
 1185 & 1302 & A2V	  & 128 & 4  \\
 1280 & 1366 & A2V	  & 102 & 4  \\
\hline
\end{tabular}
\end{table}
The \vsini\ are attributed as the mean of available values weighted by the
inverse of their variance. Trace of the membership to the different subsamples is kept 
and listed in column (5).
The composition in terms of
proportions of each subsample is represented as a pie chart in
Fig.~\ref{camembert}. The catalogue of Abt \& Morrell contributes to
the four fifths of the sample, and  the remaining fifth is composed of
new measurements derived by Fourier transforms. 

\begin{figure}[!htp]
        \centering
\resizebox{\hsize}{!}{\includegraphics{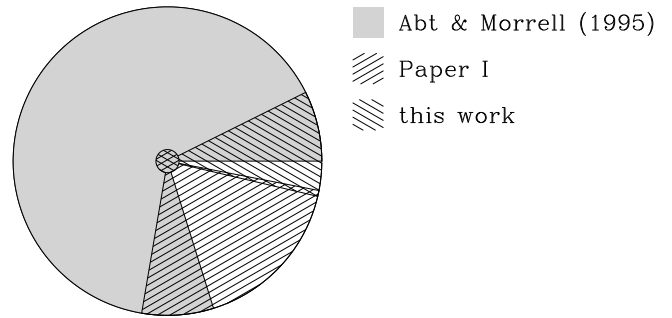}}  
        \caption{Pie chart of the subsample membership of the stars in the
          total \vsini\ sample. Multiple membership is represented by superimposed patterns.}
        \label{camembert}
\end{figure}

The total sample is displayed in Fig.~\ref{complet}.a, as a density
plot in equatorial coordinates. This distribution on the sky partly
reflects the distribution in the solar neighborhood, and the density
is slightly higher along the galactic plane (indicated by a dashed
line). Note that the cell in
 equatorial coordinates with the highest density (around $\alpha=5$~h, 
 $\delta = 23\,\degr$) in Fig.~\ref{complet}.a corresponds to the
 position of the Hyades open cluster. The lower density in the
 southern hemisphere is discussed hereafter in terms of completeness
 of the sample.

\subsection{Completeness} 
Except for a handful of stars, all belong to the HIPPARCOS catalogue. The 
latter
is complete up to a limiting magnitude $V_\mathrm{lim}$ which depends
on the galactic latitude $b$ \citep{Hip}:
\begin{equation}
\label{Vlim}
V_\mathrm{lim} = 7.9 + 1.1\,\sin|b|.
\end{equation} 

This limit $V_\mathrm{lim}$ is faint enough for counts of A-type stars
among the HIPPARCOS catalogue to allow the
estimate of the completeness of the \vsini\ sample. This sample is
north-south asymmetric because of the way it is gathered. 
Abt \& Morrell observed A-type stars from Kitt Peak, and the range of
declinations is limited from  $\delta = -30\,\degr$ to $\delta =
+70\,\degr$, and these limits can be seen in Fig.~\ref{complet}.a.
Whereas the northern part  of the sample benefits from the large number of
stars in the catalogue from Abt \& Morrell,  the southern part
 mainly comes from Paper~I.
Thus the completeness is derived for each equatorial
hemisphere. Figures~\ref{complet}.b and  \ref{complet}.c
 display the histograms in $V$ magnitude of the \vsini\ sample compared
 to the HIPPARCOS data, for $\delta > 0\,\degr$ and $\delta <
 0\,\degr$ respectively. For both sources, only the spectral interval
 from B9 to F0-type stars is taken into account. Moreover data are 
 censored, taking $V=8.0$ as the faintest magnitude.
\begin{figure*}[!htp]
        \centering
\resizebox{\hsize}{!}{\includegraphics{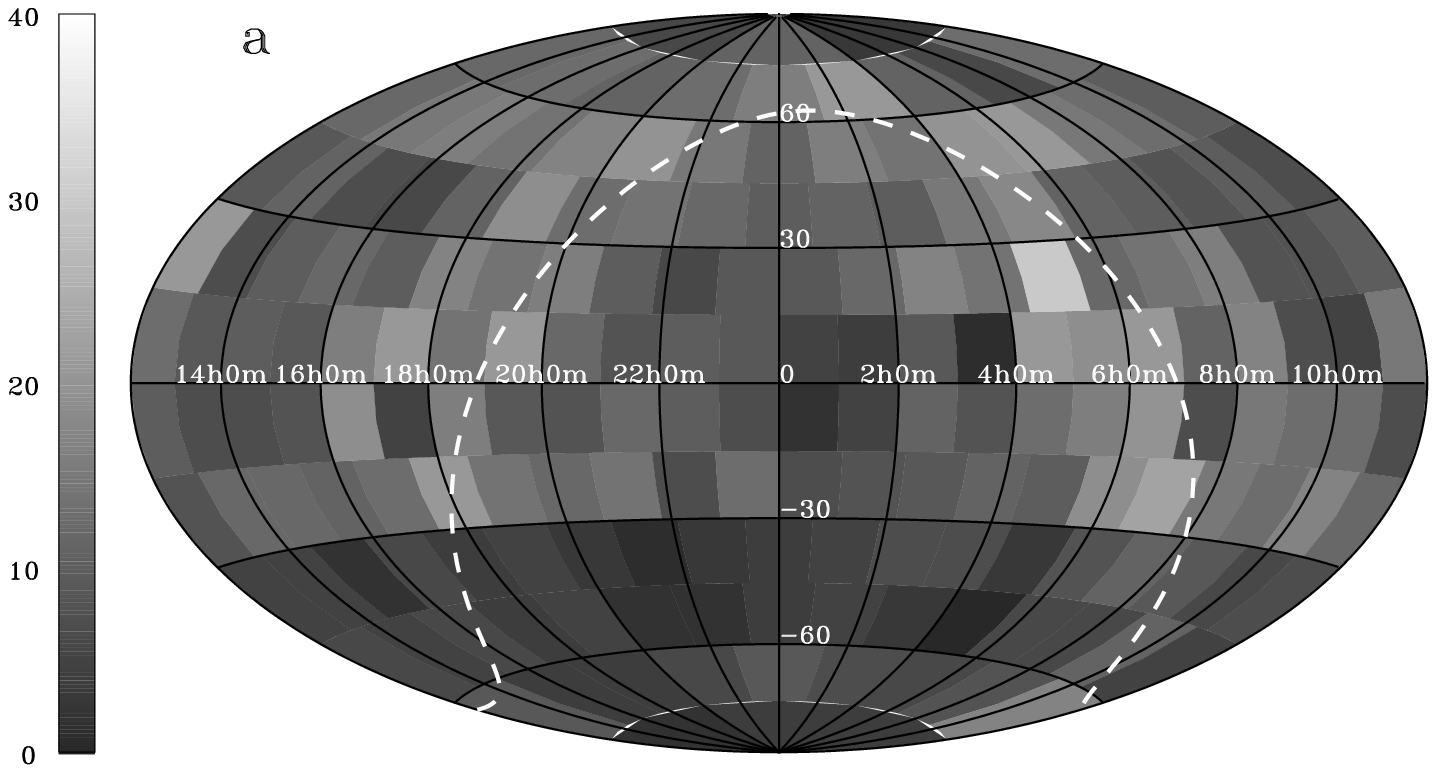}\includegraphics{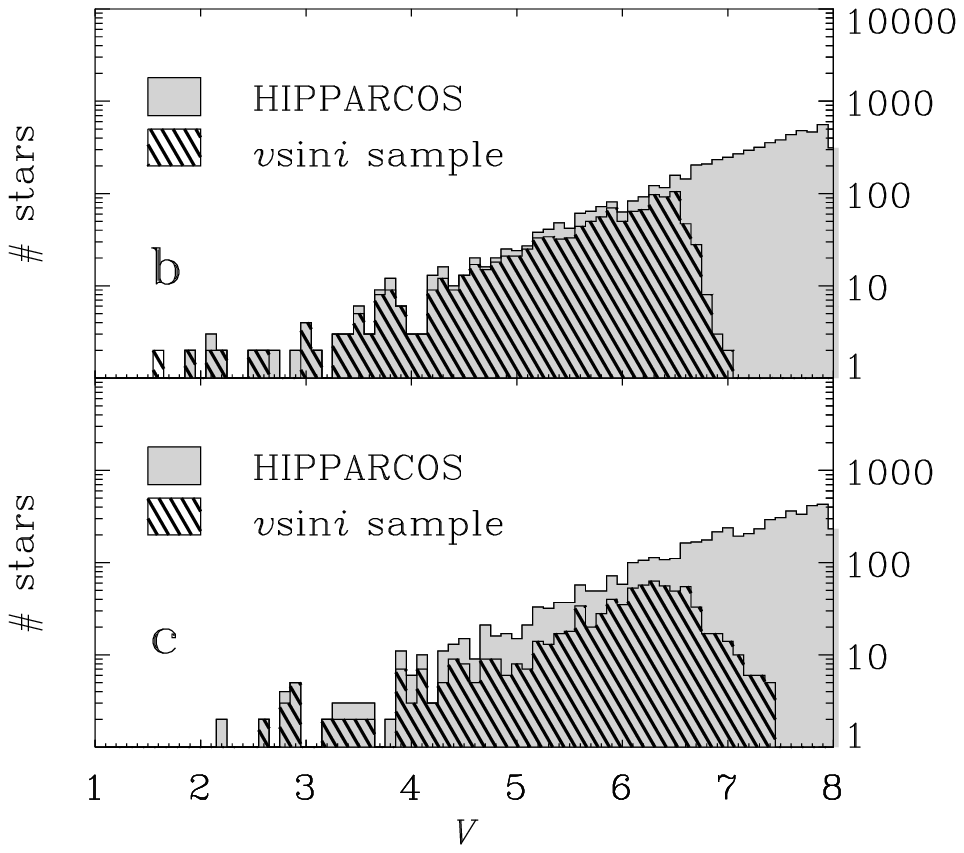}}  
        \caption{{\bf a)} Density of the \vsini\ sample on the
          sky. Counts over 15\degr$\times$15\degr\ bins in equatorial
          coordinates are indicated by the grey scale. The dashed line 
          stands for the galactic equator. {\bf b)} and
          {\bf c)} represent the counts in magnitude bins of the
          \vsini\ sample compared to the A-type stars in the HIPPARCOS 
          catalogue for the northern and southern hemisphere respectively.}
        \label{complet}
\end{figure*}

The completeness of the northern part is 80\,\% at $V=6.5$~mag. This
reflects the completeness of the Bright Star Catalogue \citep{Bsc1} from which stars 
from Abt \& Morrell are issued. In the southern part, it can be seen
that the distribution of magnitudes goes fainter, but the completeness 
is far lower and reaches 50\,\% at $V=6.5$~mag. These numbers apply
to the whole spectral range from B9 to F0-type stars, and they differ 
when considering smaller spectral bins. For the A1-type bin for
instance, the completeness reaches almost 90~\% and 70~\% for the
northern and southern hemispheres respectively, at $V=6.5$~mag.

\section{Summary and conclusions}

The determination  of projected rotational velocities is sullied with several
effects which affect the measurement. The blend of spectral lines tends to 
produce an overestimated value of \vsini, whereas the lowering of the measured
continuum level due to high rotation tends to lower the derived
\vsini. The solution lies in a good choice of candidate lines to measure the
rotational velocity. The use of the additional spectral range
4500--4600~\AA, compared to the observed domain in Paper~I, allows for the 
choice of reliable lines that can be measured even in case of high
rotational broadening and reliable anchors of the continuum, for the
considered range of spectral types.
The \vsini\ is derived from the first zero of Fourier transform of
line profiles chosen among 23 candidate lines according to the
spectral type and the rotational blending.
It gives resulting \vsini\ for 249 stars, with a precision of about 5~\%.

The systematic shift with \vsini\ standard stars from SCBWP, already
detected in Paper~I, is confirmed in this work. SCBWP's values are
underestimated, smaller by a factor of 0.8 on average, according to
common stars in the northern sample. When joining both intersections
of northern and southern samples with standard stars from SCBWP, the
relation between the two scales is about
$\vsini=1.03\,{\vsini}_\mathrm{SCBWP}+7.7$, using these 52 stars in common.
This is approximately our findings concerning the catalogue made by
\citet{AbtMol95}. They derive their \vsini\ from the calibration
built by SCBWP, and reproduce the systematic shift.

In the aim of gathering a large and homogeneous sample of  projected
rotational velocities for A-type stars, the new data, from the present
paper and from Paper~I, are merged with the catalogue of Abt \&
Morrell. First, the \vsini\ from the latter catalogue are statistically
corrected from the above mentioned systematic shift. The final sample contains
\vsini\ for 2151 B8- to F2-type stars.

The continuation of this work will consist in determining and
analyzing the
distributions of rotational velocities (equatorial and angular) for different sub-groups of spectral type,
starting from the \vsini. 

\begin{acknowledgements}
We insist on warmly thanking Dr~M.~Ramella to have
provided the programme of determination of the rotational velocities. 
We are also very grateful to Dr~R.~Faraggiana for her precious advice 
about the analysis of the spectra. We should like to acknowledge
Dr~F.~Sabati\'e for his careful reading of the manuscript.
\end{acknowledgements}

\appendix
\section{notes on stars with uncertain rotational velocity}
\label{Notes0}
\subsection{Stars with no selected line} 
In a few cases, the selected lines are all discarded either from their 
Fourier profile or from their skewness (Table~\ref{skewness}). For
these stars, an uncertain value of \vsini\ is derived from the lines
that should have been discarded. They are indicated by a colon and
flagged as ``NO'' in Table~\ref{results}. These objects are listed below. It is 
worth noticing that none of them have spectra collected in $\Lambda_3$ 
spectral range:
\begin{itemize} 
 \item  \object{HD~27459} and \object{HD~27819} are early-F and late-A  
   type stars from Hyades open cluster. Due to their rotational
   broadening and their late spectral type, no lines are retained in
   $\Lambda_1$. Their uncertain \vsini\ are respectively 78: and
   41:~\kms. \citet{VaeMor99} measure their rotational velocity and
   find 78 and 47~\kms.

 \item  \object{HD~32537} is a early-F type star, whose \vsini\ derived 
   from discarded line in $\Lambda_1$ is 27:~\kms. The rotational velocity
   measured by \citet{Hua00} is 21~\kms.
       
 \item  \object{HD~53929} is a late-B giant star, its spectra in
   $\Lambda_1$ exhibit very few lines. The \vsini\ is derived from
   \ion{Fe}{ii} 4233 that is usually discarded because blended. For
   this spectral type, blend of the iron line is less pronounced, and
   the \vsini\ is 25:~\kms. \citet{Dwy_98} find the same value.
   
 \item  Following stars are high rotators whose only measurable line in $\Lambda_1$
   is the blended line \ion{Fe}{ii} 4233. 
   \begin{itemize} 
   \item  \object{HD~59059},   260:~\kms 
   \item  \object{HD~77104},   183:~\kms    
   \item  \object{HD~111469},  197:~\kms
   \item  \object{HD~224404},  190:~\kms
   \end{itemize}
\end{itemize}

\subsection{Stars with high external error}
\label{Notes1}
A few stars of the sample exhibit an external error higher than the
estimation carried on in Sect.~\ref{precision}. It can be the
signature of a multiple system.
The following stars have variable \vsini\ from spectrum to spectrum
and are labeled as ``SS'' in Table~\ref{results}:
\begin{itemize}
\item      \object{HD~4321} has a mean $\vsini =25\pm 4$~\kms, but
  for its different spectra, this value is respectively   38, $22\pm
  3$ and $25\pm 1$~\kms.
\item     \object{HD~40446} has a mean $\vsini =27\pm 5$~\kms\
  (respectively 27, 32 and 22~\kms). The HIPPARCOS solution for this
  object is flagged ``suspected non single''.
\item     \object{HD~40626} has a mean    $\vsini =18\pm
  3$~\kms\   ($16\pm 1$ and $20\pm 1$~\kms). The \vsini\ found by
  \citet{Raa_89} is 10.8~\kms.
\item    \object{HD~103578}  has a mean    $\vsini =16\pm 4$~\kms\  ($14\pm 1$, $19\pm
  1$~\kms). It is a SB2 system, and \citet{Hul_01} who study its orbital
  motion, find $11.03\pm 0.14$~\kms\ and $8.12\pm 0.27$~\kms\ for the
  \vsini\ of each component.
\end{itemize}


\section{notes on {\boldmath\vsini} standard stars with discrepant rotational velocity}
\label{Notes3}
The common stars, among SCBWP's data and this sample, which 
exhibit the largest differences in \vsini\ between both studies, are
listed in Table~\ref{vsini_litt}. They are detailed below.

\begin{itemize}
\item \object{Alhena} ($\gamma$~Gem, HD~47105) is a single-lined
  spectroscopic binary with a solar-type companion \citep{Scz_97}.
  \citet{Hil95} carries out spectral synthesis on some A-type stars in
  the aim of deriving abundances and finds $\vsini=11.2$~\kms\ for
  $\gamma$~Gem.  Scholz et al. measure the velocity using Fourier
  techniques on three spectral lines around 6\,150~\AA\ and derived
  $\vsini=10.2\,\kms\,\pm 0.2\,\kms$. These two measurements are close
  to the value given by SCBWP: $<10$~\kms.
  
\item \object{30~Mon} (HD~71155) has no determination of \vsini\ in
  the literature, independent from Slettebak's systems. It is
  nevertheless a common star between this work and Paper~I, as
  described in Sect.~\ref{so-no}. The ``probable binarity'' of 30~Mon,
  detected by \citet{Grr_99a} could explain the discrepancy in
  \vsini. It is worth noticing that discrepancy is also observed with
  the \vsini\ determined in Paper~I for this same object (Table~\ref{vsini_NS}).
  
\item \object{Merak} ($\beta$~UMa, HD~95418) is known as a
  ``Vega-type'' star, supposed to be surrounded by protoplanetary
  material \citep{Aun85}. Hill derives its \vsini\ as a sub-product of
  its abundance analysis. The value of 44.8~\kms\ is significantly
  larger than the one found by SCBWP and is well in agreement with
  ours.
  
\item \object{$\theta$~Leo} (HD~97633) is studied by \citet{Fel98},
  \citet{Lee89} and \citet{Hil95}, who give \vsini\ quite in agreement
  with our derived value, as already pointed out in Paper~I.
  \citet{SmhDwy93} apply Fourier method to IUE spectra to derive
  \vsini\ for B and early A-type stars. Their finding for $\theta$~Leo
  is 24~\kms.
  
\item \object{Phecda} ($\gamma$~UMa, HD~103287) is a spectroscopic
  binary. \citet{Gry80b} uses FT of the \ion{Mg}{ii} line profile and
  derives $\vsini=154\,\pm 4\,\kms$.
  
\item \object{Thuban} ($\alpha$~Dra, HD~123299) is a well-known
  single-lined spectroscopic binary. \citet{LenScz93} use Bessel
  functions on the profile of \ion{Mg}{ii} 4481 and derive a
  $\vsini=27$~\kms\ almost identical to the value found in this paper.
  \citet{Adn_87} derive a lower velocity, but still larger than
  15~\kms\ given by SCBWP.
  
\item \object{$\sigma$~Boo} (HD~128167) is a slowly rotating F-type
  star. Fitting the line profiles with synthetic ones, \citet{Som82}
  derives $\vsini = 7.5$~\kms. \citet{Gry84}, using Fourier method,
  finds an identical value. From the width of the CORAVEL
  cross-correlation function, \citet{BezMar84} derive $\vsini =
  8.1\,\pm 0.8$~\kms\ and from the width of the spectral lines,
  \citet{Fel97} finds 7.8~\kms.
  
\item \object{Alphecca} ($\alpha$~CrB, HD~139006) is an Algol-type
  eclipsing binary. The primary component is an A0V-type star and
  its projected rotational velocity is estimated at 110~\kms\ by
  SCBWP. \citet{Gry80b} finds $\vsini=127\,\kms\,\pm 4\,\kms$ from
  Fourier analysis.  
  
\item \object{$\tau$~Her} (HD~147394) is a slowly pulsating B star
  whose period derived from HIPPARCOS data is $P=1.25$~d.
  \citet{MaaHia00} detect the line-profile variation by studying
  \ion{He}{i}~4471 and \ion{Mg}{ii}~4481. These variations could imply
  changes in the derived rotational velocity as both SCBWP's and ours
  rely partly on \ion{Mg}{ii}~4481. \citet{SmhDwy93} derive, using
  Fourier method, $\vsini = 32$~\kms.
  
\item \object{Vega} ($\alpha$~Lyr, HD~172167) used to serve as a zero
  rotation standard, when limited spectral resolution prevented
  detection of low \vsini. The rotational velocity given by SCBWP is
  $< 10$~\kms. \citet{Gry80a} disagrees with this ``distinctly too
  small'' value and finds with Fourier techniques a larger value:
  23.4~\kms\ $\pm 0.4$~\kms. Vega is now suspected to be a rapid
  rotator nearly seen pole-on, and \citet{Hil95}, \citet{Gur_94} and \citet{ErrNoh02}
  using synthetic spectra, derive its \vsini\ and respectively give
  22.4, 21.8~\kms\ $\pm 0.2$~\kms, and 23.2~\kms.
  
\item \object{$\gamma$~Lyr} (HD~176437) has no indication of \vsini\
  determination or spectral peculiarity in the literature.
  
\item \object{$\epsilon$~Aqr} (HD~198001) has a $\vsini=85$~\kms\ 
  according to SCBWP, smaller than the value in this work. As already
  mentioned in Sect.~\ref{so-no}, this star has not been observed in
  $\Lambda_3$ spectral range, so that the derived \vsini\ is probably
  overestimated. Nevertheless, $\epsilon$~Aqr also belongs to the
  sample measured in Paper~I and was already found discrepant, with a
  \vsini\ rather in agreement with those derived by Hill and \citet{Dun_97}.
\end{itemize}

\section{notes on common stars with \citet{AbtMol95} rejected by sigma-clipping}
\label{Notes4}
When merging the sample from  \citet{AbtMol95} with the new
measurements using Fourier transforms, common data are compared in
order to compute the scaling law between both samples. Aberrant points 
are discarded using a sigma-clipping algorithm. These stars are listed 
and discussed, and their \vsini\ are indicated in \kms\ ($\vsini_{\mathrm{I}\cup
    \mathrm{II}} / \vsini_{\mathrm{AM}}$):

\begin{itemize}
\item \object{HD~18778}   (26 / 43) is a spectroscopic binary whose
  period is measured by \citet{Abt61}: $P_\mathrm{orb} = 11\fd 665$. 
  Its \vsini\ is found to be 20~\kms\ by \citet{Dov85}.
\item       \object{HD~27962} (11 / 15) is a binary star from
  HIPPARCOS data. 
\item       \object{HD~29573} (26 / 31) is detected as variable from
  HIPPARCOS observations, with unsolved solution.
\item       \object{HD~32115} (12 / 15). 
\item       \object{HD~33204} (160 / 45) is an Am star in Hyades and
  is the primary component of the triple system ADS 3730. It is detected as a binary star by
  HIPPARCOS. \citet{Dei_00} compute the orbital parameters of this SB1 
  star, and derive its $\vsini=24.0\pm 2.4$~\kms.
\item       \object{HD~37594} (14 / 15)$^\star$. 
\item       \object{HD~43760} (9 / 15) is suspected to be variable in
  radial velocity \citep{Hua00}, and its $\vsini$ is found to be 13~\kms.  
\item       \object{HD~50644} (12 / 13)$^\star$ is indicated as
  micro-variable in the HIPPARCOS catalogue.   Speckle observations \citep{Haf_97}
  allow the measurement of the separation of the system components
  $\rho= 0\farcs 159$.
\item       \object{HD~71297} (11 / 13)$^\star$ is a suspected binary
  from HIPPARCOS data.  
\item       \object{HD~72660} (9 / 10)$^\star$ is a low rotator $\vsini=6.5$~\kms\ in \citet{NinWan00}
and 6~\kms\ in \citet{Vae99}. 
\item      \object{HD~111786} (45 / 135) is a periodic variable star
  in the HIPPARCOS catalogue.  Its spectrum is composite
  \citep{Faa_01} and HD~111786 appears to be a multiple system.
\item      \object{HD~115604} (15 / 15)$^\star$.  
\item \object{HD~132145}   (15 / 30) has also been observed by \citet{Raa_89} who
  find $\vsini= 16.6$~\kms.
\item \object{HD~158716}   (15 / 35) has also been observed by \citet{Raa_89} who
  find $\vsini= 19.6$~\kms.
\item  \object{HD~159082}   (21 / 30) has also been observed by \citet{Raa_89} who
  find $\vsini=  19.6$~\kms.
  SB $P_\mathrm{orb} = 6.79750$ \citep{StdWey84}
\item \object{HD~159480}   (28 / 40) is analyzed by
  \citet{Lee89} who derives a \vsini\ from spectral synthesis: 27~\kms. HD~159480 has a negative result in speckle interferometry \citep{Mcr_87}.
\item \object{HD~159834}   (18 / 28).
\item      \object{HD~160839} (40 / 51).
\item      \object{HD~175687} (10 / 10)$^\star$.
\item \object{HD~187340}   (15 / 33).
\item      \object{HD~192640} (86 / 35) is an unsolved variable star
  in the HIPPARCOS catalogue.
\item      \object{HD~203858} (14 / 70) is a spectroscopic
  binary. \citet{Mia_93}, using speckle interferometry, estimate the separation between 
  components $\rho<0\farcs 039$.
\item \object{HD~208108}   (28 / 35) has also been observed by \citet{Raa_89} who
  find $\vsini=  29.0$~\kms. 
\end{itemize}
$^\star$: these stars show very coherent values between both samples,
but these low \vsini\ are considered ``discrepant'' with respect to
the scaling law (Eq.~\ref{scale}).



\begin{thebibliography}{53}
\expandafter\ifx\csname natexlab\endcsname\relax\def\natexlab#1{#1}\fi

\bibitem[{{Abt}(1961)}]{Abt61}
{Abt}, H.~A. 1961, ApJS, 6, 37

\bibitem[{{Abt} \& {Morrell}(1995)}]{AbtMol95}
{Abt}, H.~A. \& {Morrell}, N.~I. 1995, ApJS, 99, 135

\bibitem[{{Adelman} {et~al.}(1987){Adelman}, {Bolcal}, {Kocer}, \&
  {Inelmen}}]{Adn_87}
{Adelman}, S.~J., {Bolcal}, C., {Kocer}, D., \& {Inelmen}, E. 1987, PASP, 99,
  130

\bibitem[{{Aumann}(1985)}]{Aun85}
{Aumann}, H.~H. 1985, PASP, 97, 885

\bibitem[{{Batten} {et~al.}(1989){Batten}, {Fletcher}, \& {MacCarthy}}]{Ban_89}
{Batten}, A.~H., {Fletcher}, J.~M., \& {MacCarthy}, D.~G. 1989, 8th Catalogue
  of the orbital elements of spectroscopic binary systems (Victoria: Dominion
  Astrophysical Observatory)

\bibitem[{{Benz} \& {Mayor}(1984)}]{BezMar84}
{Benz}, W. \& {Mayor}, M. 1984, A\&A, 138, 183

\bibitem[{{Carroll}(1933)}]{Cal33}
{Carroll}, J.~A. 1933, MNRAS, 93, 478

\bibitem[{{Debernardi} {et~al.}(2000){Debernardi}, {Mermilliod}, {Carquillat},
  \& {Ginestet}}]{Dei_00}
{Debernardi}, Y., {Mermilliod}, J.-C., {Carquillat}, J.-M., \& {Ginestet}, N.
  2000, A\&A, 354, 881

\bibitem[{{Dobrichev}(1985)}]{Dov85}
{Dobrichev}, V. 1985, Astrofizicheskie Issledovaniya Sofia, 4, 40

\bibitem[{{Dunkin} {et~al.}(1997){Dunkin}, {Barlow}, \& {Ryan}}]{Dun_97}
{Dunkin}, S.~K., {Barlow}, M.~J., \& {Ryan}, S.~G. 1997, MNRAS, 286, 604

\bibitem[{{Dworetsky} {et~al.}(1998){Dworetsky}, {Jomaron}, \&
  {Smith}}]{Dwy_98}
{Dworetsky}, M.~M., {Jomaron}, C.~M., \& {Smith}, C.~A. 1998, A\&A, 333, 665

\bibitem[{{Erspamer} \& {North}(2002)}]{ErrNoh02}
{Erspamer}, D. \& {North}, P. 2002, A\&A, 383, 227

\bibitem[{ESA(1997)}]{Hip}
ESA. 1997, The Hipparcos and Tycho Catalogues, ESA-SP 1200

\bibitem[{{Faraggiana} {et~al.}(2001){Faraggiana}, {Gerbaldi}, {Bonifacio}, \&
  {Fran{\c c}ois}}]{Faa_01}
{Faraggiana}, R., {Gerbaldi}, M., {Bonifacio}, P., \& {Fran{\c c}ois}, P. 2001,
  A\&A, 376, 586

\bibitem[{{Fekel}(1997)}]{Fel97}
{Fekel}, F.~C. 1997, PASP, 109, 514

\bibitem[{{Fekel}(1998)}]{Fel98}
{Fekel}, F.~C. 1998, in Precise stellar radial velocities, IAU Colloquium 170,
  ed. J.~B. {Hearnshaw} \& C.~D. {Scarfe}, E64

\bibitem[{{Gillet} {et~al.}(1994){Gillet}, {Burnage}, {Kohler}, {Lacroix},
  {Adrianzyk}, {Baietto}, {Berger}, {Goillandeau}, {Guillaume}, {Joly},
  {Meunier}, {Rimbaud}, \& {Vin}}]{Git_94}
{Gillet}, D., {Burnage}, R., {Kohler}, D., {et~al.} 1994, A\&AS, 108, 181

\bibitem[{{Gray}(1980{\natexlab{a}})}]{Gry80b}
{Gray}, D.~F. 1980{\natexlab{a}}, PASP, 92, 771

\bibitem[{{Gray}(1980{\natexlab{b}})}]{Gry80a}
---. 1980{\natexlab{b}}, PASP, 92, 154

\bibitem[{{Gray}(1984)}]{Gry84}
---. 1984, ApJ, 281, 719

\bibitem[{{Grenier} \& {Burnage}(1995)}]{GrrBue95}
{Grenier}, S. \& {Burnage}, R. 1995, in De l'utilisation des donn\'ees
  Hipparcos, ed. M.~{Fr\oe schl\'e} \& F.~{Mignard}, GDR 051, 177

\bibitem[{{Grenier} {et~al.}(1999){Grenier}, {Burnage}, {Faraggiana},
  {Gerbaldi}, {Delmas}, {G\'omez}, {Sabas}, \& {Sharif}}]{Grr_99a}
{Grenier}, S., {Burnage}, R., {Faraggiana}, R., {et~al.} 1999, A\&AS, 135, 503

\bibitem[{{Gulliver} {et~al.}(1994){Gulliver}, {Hill}, \& {Adelman}}]{Gur_94}
{Gulliver}, A.~F., {Hill}, G., \& {Adelman}, S.~J. 1994, ApJ Lett., 429, L81

\bibitem[{{Hartkopf} {et~al.}(2000){Hartkopf}, {Mason}, {McAlister}, {Roberts},
  {Turner}, {ten Brummelaar}, {Prieto}, {Ling}, \& {Franz}}]{Haf_00}
{Hartkopf}, W.~I., {Mason}, B.~D., {McAlister}, H.~A., {et~al.} 2000, AJ, 119,
  3084

\bibitem[{{Hartkopf} {et~al.}(1997){Hartkopf}, {McAlister}, {Mason},
  {Brummelaar}, {Roberts}, {Turner}, \& {Wilson}}]{Haf_97}
{Hartkopf}, W.~I., {McAlister}, H.~A., {Mason}, B.~D., {et~al.} 1997, AJ, 114,
  1639

\bibitem[{{Hill}(1995)}]{Hil95}
{Hill}, G.~M. 1995, A\&A, 294, 536

\bibitem[{{Hoffleit} \& {Jaschek}(1982)}]{Bsc1}
{Hoffleit}, D. \& {Jaschek}, C. 1982, The Bright Star Catalogue, 4th edn. (New
  Haven, Conn.: Yale University Observatory)

\bibitem[{{Holweger} {et~al.}(1999){Holweger}, {Hempel}, \& {Kamp}}]{Hor_99}
{Holweger}, H., {Hempel}, M., \& {Kamp}, I. 1999, A\&A, 350, 603

\bibitem[{{Horn} {et~al.}(1996){Horn}, {Kub\'at}, {Harmanec}, {Koubsk\'y},
  {Hadrava}, {\v{S}imon}, {\v{S}tefl}, \& {\v{S}koda}}]{Hon_96}
{Horn}, J., {Kub\'at}, J., {Harmanec}, P., {et~al.} 1996, A\&A, 309, 521

\bibitem[{{Hui-Bon-Hoa}(2000)}]{Hua00}
{Hui-Bon-Hoa}, A. 2000, A\&AS, 144, 203

\bibitem[{{Hui-Bon-Hoa} \& {Alecian}(1998)}]{HuaAln98}
{Hui-Bon-Hoa}, A. \& {Alecian}, G. 1998, A\&A, 332, 224

\bibitem[{{Hummel} {et~al.}(2001){Hummel}, {Carquillat}, {Ginestet}, {Griffin},
  {Boden}, {Hajian}, {Mozurkewich}, \& {Nordgren}}]{Hul_01}
{Hummel}, C.~A., {Carquillat}, J.-M., {Ginestet}, N., {et~al.} 2001, AJ, 121,
  1623

\bibitem[{{Jefferys} {et~al.}(1998{\natexlab{a}}){Jefferys}, {Fitzpatrick}, \&
  {McArthur}}]{Jes_98a}
{Jefferys}, W.~H., {Fitzpatrick}, M.~J., \& {McArthur}, B.~E.
  1998{\natexlab{a}}, Celest. Mech., 41, 39

\bibitem[{{Jefferys} {et~al.}(1998{\natexlab{b}}){Jefferys}, {Fitzpatrick},
  {McArthur}, \& {McCartney}}]{Jes_98b}
{Jefferys}, W.~H., {Fitzpatrick}, M.~J., {McArthur}, B.~E., \& {McCartney},
  J.~E. 1998{\natexlab{b}}, GaussFit: A System for least squares and robust
  estimation, User's Manual, Dept. of Astronomy and McDonald Observatory,
  Austin, Texas

\bibitem[{{Johansen} \& {S\o rensen}(1997)}]{JonSon97}
{Johansen}, K.~T. \& {S\o rensen}, H. 1997, Inf. Bull. Variable Stars, 4533, 1

\bibitem[{{Kurucz}(1993)}]{Kuz93}
{Kurucz}, R.~L. 1993 (Kurucz CD-ROM, Cambridge, Smithsonian Astrophysical
  Observatory)

\bibitem[{{Lehmann} \& {Scholz}(1993)}]{LenScz93}
{Lehmann}, H. \& {Scholz}, G. 1993, in ASP Conf. Ser. 44: IAU Colloq. 138:
  Peculiar versus Normal Phenomena in A-type and Related Stars, 612

\bibitem[{{Lemke}(1989)}]{Lee89}
{Lemke}, M. 1989, A\&A, 225, 125

\bibitem[{{Liu} {et~al.}(1997){Liu}, {Gies}, {Xiong}, {Riddle}, {Bagnuolo},
  {Barry}, {Ferrara}, {Hartkopf}, {Hooda}, {Mason}, {McAlister}, {Roberts}, \&
  {Sowers}}]{Liu_97}
{Liu}, N., {Gies}, D.~R., {Xiong}, Y., {et~al.} 1997, ApJ, 485, 350

\bibitem[{{Masuda} \& {Hirata}(2000)}]{MaaHia00}
{Masuda}, S. \& {Hirata}, R. 2000, A\&A, 356, 209

\bibitem[{{McAlister} {et~al.}(1987){McAlister}, {Hartkopf}, {Hutter}, {Shara},
  \& {Franz}}]{Mcr_87}
{McAlister}, H.~A., {Hartkopf}, W.~I., {Hutter}, D.~J., {Shara}, M.~M., \&
  {Franz}, O.~G. 1987, AJ, 93, 183

\bibitem[{{Miura} {et~al.}(1993){Miura}, {Ni-Ino}, {Baba}, {Iribe}, \&
  {Isobe}}]{Mia_93}
{Miura}, N., {Ni-Ino}, M., {Baba}, N., {Iribe}, T., \& {Isobe}, S. 1993, Publ.
  Natl. Astron. Obs. Jpn., 3, 153

\bibitem[{{Nielsen} \& {Wahlgren}(2000)}]{NinWan00}
{Nielsen}, K. \& {Wahlgren}, G.~M. 2000, A\&A, 356, 146

\bibitem[{{Nordstr\"om} \& {Johansen}(1994)}]{NomJon94}
{Nordstr\"om}, B. \& {Johansen}, K.~T. 1994, A\&A, 291, 777

\bibitem[{{Ramella} {et~al.}(1989){Ramella}, {B\"ohm}, {Gerbaldi}, \&
  {Faraggiana}}]{Raa_89}
{Ramella}, M., {B\"ohm}, C., {Gerbaldi}, M., \& {Faraggiana}, R. 1989, A\&A,
  209, 233

\bibitem[{{Royer} {et~al.}(2002){Royer}, {Gerbaldi}, {Faraggiana}, \&
  {G\'omez}}]{Ror_02a}
{Royer}, F., {Gerbaldi}, M., {Faraggiana}, R., \& {G\'omez}, A.~E. 2002, A\&A,
  381, 105, (Paper I)

\bibitem[{{Scholz} {et~al.}(1997){Scholz}, {Lehmann}, {Harmanec}, {Gerth}, \&
  {Hildebrandt}}]{Scz_97}
{Scholz}, G., {Lehmann}, H., {Harmanec}, P., {Gerth}, E., \& {Hildebrandt}, G.
  1997, A\&A, 320, 791

\bibitem[{{Slettebak} {et~al.}(1975){Slettebak}, {Collins}, {Boyce}, {White},
  \& {Parkinson}}]{Slk_75}
{Slettebak}, A., {Collins}, I. G.~W., {Boyce}, P.~B., {White}, N.~M., \&
  {Parkinson}, T.~D. 1975, ApJS, 29, 137, (SCBWP)

\bibitem[{{Smith} \& {Dworetsky}(1993)}]{SmhDwy93}
{Smith}, K.~C. \& {Dworetsky}, M.~M. 1993, A\&A, 274, 335

\bibitem[{{Soderblom}(1982)}]{Som82}
{Soderblom}, D.~R. 1982, ApJ, 263, 239

\bibitem[{{Stickland} \& {Weatherby}(1984)}]{StdWey84}
{Stickland}, D.~J. \& {Weatherby}, J. 1984, A\&AS, 57, 55

\bibitem[{{Varenne}(1999)}]{Vae99}
{Varenne}, O. 1999, A\&A, 341, 233

\bibitem[{{Varenne} \& {Monier}(1999)}]{VaeMor99}
{Varenne}, O. \& {Monier}, R. 1999, A\&A, 351, 247

\end{thebibliography}
\end{document}